\documentclass{elsart}

\usepackage[dvips]{graphicx}
\usepackage{natbib}

\journal{Annals of Physics}

\begin{document}
\begin{frontmatter}

\title{SUSY-hierarchy of one-dimensional reflectionless
potentials\thanksref{www.1}}

\thanks[www.1]{The paper is available on-line in
http://dx.doi.org/10.1016/j.aop.2004.11.004}

\author{Sergei~P.~Maydanyuk}
\ead{maidan@kinr.kiev.ua}

\address{
Institute for Nuclear Research,
National Academy of Sciences of Ukraine \\
prosp. Nauki, 47, Kiev-28, 03680, Ukraine}

% \date{\today}

\begin{abstract}
A class of one-dimensional reflectionless potentials is studied.
It is found, that all possible types of the reflectionless
potentials can be combined into one SUSY-hierarchy with a
constant potential.
An approach for determination of a general form of the reflectionless
potential on the basis of construction of such a hierarchy by the
recurrent method is proposed.
A general integral form of interdependence between superpotentials
with neighboring numbers of this hierarchy, opening a possibility to
find new reflectionless potentials, is found and has a simple
analytical view.
It is supposed that any possible type of the reflectionless potential
can be expressed through finite number of elementary functions
(unlike some presentations of the reflectionless potentials, which
are constructed on the basis of soliton solutions or are shape
invariant in one or many steps with involving scaling of parameters,
and are expressed through series).
An analysis of absolute transparency existence for the potential
which has the inverse power dependence on space coordinate (and
here tunneling is possible), i.~e. which has the form
$V(x) = \pm \alpha / |x-x_{0}|^{n}$
(where $\alpha$ and $x_{0}$ are constants, $n$ is natural number),
is fulfilled. It is shown that such a potential can be
reflectionless at $n = 2$ only.
A SUSY-hierarchy of the inverse power reflectionless potentials
is constructed. Isospectral expansions of this hierarchy is
analyzed.
%
% A class of one-dimensional reflectionless potentials is studied.
% It is found, that all possible types of the reflectionless
% potentials can be combined into one SUSY-hierarchy with a
% constant potential.
% A recurrent approach for construction of such hierarhy is analyzed.
% A general integral form of the reflectionless superpotential is
% found.
% A SUSY-hierarchy of the inverse power reflectionless potentials
% (here tunneling is possible) is constructed.
% Isospectral expansions of this hierarchy is analyzed.
% It is supposed that any possible type of the reflectionless potential
% has a simple analytical presentation and is expressed through finite
% number of elementary functions (without use of series).
\end{abstract}

\begin{keyword}
supersymmetry, quantum mechanics \sep
exactly solvable model \sep 
reflectionless potentials \sep 
inverse power potentials \sep 
isospectral potentials \sep 
SUSY-hierarchy

\PACS
11.30.Pb \sep        % Supersymmetry
03.65.-w \sep        % Quantum mechanics
12.60.Jv \sep        % Supersymmetric models
03.65.Xp \sep        % Tunneling, traversal time, quantum Zeno dynamics
03.65.Fd             % Algebraic methods
\end{keyword}

\end{frontmatter}
%---------------------------------------------------------------------------

%---------------------------------------------------------------------------
% \maketitle
%
% \newpage
% \tableofcontents
% ***************************************************************************

% ***************************************************************************
% \newpage
\section{Introduction
\label{sec.1}}

The tunneling effect had considered else at the time of formulation
of main principles of quantum mechanics and has been studied many
years. Nevertheless, a lot of papers down to last days are
directed to research of some its properties, which look rather
unusual from the point of view of common sense (for example,
resonant tunneling, reinforcement of penetrability of a barrier
and violation of tunneling symmetry in opposite directions in the
propagation of a system of several particles, absolute transparency
for sub-barrier energies and reflection for above-barrier energies).

Original methods are developed with a purpose to understand such
properties more deeply. Here, methods of supersymmetric quantum
mechanics (SUSY QM) allow to find types of quantum systems (both in
the region of continuous energy spectrum, and in discrete one),
which potentials have penetrability coefficient equal to one for
particle propagation through them both at separate levels,
and in continuous range of the energy spectra. In the first case
one can speak about \emph{resonant tunneling}. In the second
case the quantum systems and their potentials are named as
\emph{reflectionless} or \emph{absolutely transparent}
\cite{Zakhariev.1993.PHLTA}.

A phenomenon of the resonant tunneling and, especially, papers
directed to study of its demonstration in the concrete physical
tasks, cause a significant interest \cite{Saito.1994.JCOME}. But
the reflectionless potentials having the penetrability coefficient,
practically equal to one in a whole range of the continuous
energy spectrum, are more unusual and, apparently, have other
reasons of their existence.
The reflectionless potentials can be investigated by methods of
direct and inverse problems. Here, we should like to note both the
monograph \cite{Chadan.1995}, and the excellent reviews
\cite{Zakhariev.1994.PEPAN,Zakhariev.1999.PEPAN}, where both
methods for detailed study of the properties of the reflectionless
one- and multichannel quantum systems (mainly, in the region of the
discrete energy spectrum), and approaches simple enough for their
qualitative understanding are presented. All these methods have found
their application in the scattering theory (both in the direct
problem, and in the inverse one).

On the other side, the SUSY QM methods propose own approach for
studying the properties of the reflectionless potentials. A
considerable contribution is made by the papers, where the
shape invariant potentials in one and many steps with involving
scaling of parameters or their other transformations are studied
(for example, see~\cite{Khare.1993.JPAGB,Balantekin.1997.PHRVA}). As
a separate direction one can note the papers on study of
self-similar potentials of Shabat and Spiridonov
\cite{Shabat.1992.INPEE,Spiridonov.1992.PRLTA,Barclay.1993.PHRVA},
concerned with $q$-supersymmetry \cite{Spiridonov.1992.MPLAE}.
The review \cite{Cooper.1995.PRPLC} is the best (in my opinion).

Note, that the supersymmetric methods are less developed for study
of the properties of the quantum systems in the region of the
continuous energy spectrum (here, because of different boundary
conditions, normalization conditions the interdependences between
wave functions, energy spectra for the SUSY-partners systems can be
qualitatively differed from similar interdependences of such
spectral characteristics in the region of the discrete energy
spectra).
The different classes of the reflectionless potentials are opened
by use of the different methods and a question about the account of
all possible types of the reflectionless potentials (and, probably,
about their classification) is appeared.
A lot of known reflectionless potentials is concerned with soliton
solutions in the inverse problem approach (for example, in the
monography \cite{Zakhariev.1985} they are introduced as soliton
solutions), and also with the soliton solutions in the SUSY QM
approach (one- and many steps solutions, self-similar potentails of
Shabat and Spiridonov, Rosen-Morse potentials with their
$q$-deformation), where the absolute transparency is concerned with
the above-barrier propagation, whereas considerably more rare is
the reflectionless potential with a barrier above asymptotic tails,
where the absolute transparency is observed in the sub-barrier
tunneling.
Many of known reflectionless potentials are expressed with use of
series of a rather complicated form, and any found reflectionless
potential in a simple analytical form can be useful by its
clearness for the qualitative analysis of the properties of the
reflectionless quantum systems.

In this paper we propose a way for determination of a general form
of the one-dimensional reflectionless potential on the basis of
construction of SUSY-hierarchy, which contains a constant potential.
In obtaining of the hierarchy we try to take into account all
possible types of the potentials, which belong to it. In the
construction of the hierarchy we use a recurrent way, trying to not
involve an
analysis of an obvious form of wave functions (because of an
influence of the boundary conditions, the normalization conditions
on them can be essential). We suppose, that knowledge of such a
hierarchy will allow to determine all possible types of the
reflectionless potentials, and in such a case \emph{one can explain
a nature of the absolute transparency of any reflectionless
potential by its SUSY-connection with the constant potential
through the considered SUSY-hierarchy}. Such a definition of the
reflectionless potentials looks more expanded, than the cases of the
reflectionless soliton potentials (for example,
see~\cite{Zakhariev.1985}) or the shape invariant reflectionless
potentials.
       
In the construction of the considered SUSY-hierarchy all found by
us reflectionless potentials have simple analytical forms and are
expressed through elementary functions. According to the analysis,
the complete SUSY-hierarchy includes both the reflectionless
potentials constructed on a basis of the soliton solutions, and
a set of the shape invariant reflectionless potentials,
many from which are expressed with use of series. It
results us to the idea that any reflectionless potential can be
presented in a simple analytical form through finite number
of elementary functions, that for any reflectionless potential
with a known complicated representation in form of series
there is another analytical representation in a simple form with use
of elementary functions. However, one can make a final conclusion
after a detailed analysis of all possible types of the
reflectionless potentials (that we leave for future research).
The simple evident representation of the reflectionless potentials
is useful, because it, undoubtedly, will result to easier study of
such potentials and, as result, to deeper understanding of the
resonant tunneling, the absolute transparency and other unusual
properties of tunneling.

We continue an analysis (started in \cite{Maydanyuk.2004.PAST.refl})
of the quantum systems with a completely continuous energy spectrum,
which potentials have the inverse power dependence on space
coordinate (here, the absolute transparency is possible during
tunneling).
We propose an approach for construction of the SUSY-hierarchy
of the reflectionless inverse power potentials (obtained at the
first time), and also we construct other special cases of
SUSY-hierarchies of the reflectionless potentials.

For the first time we obtain some types of the reflectionless
potentials, which have a simple analytical form, are expressed
with use of the elementary functions and (at $x>0$)
qualitatively remind a radial potential of interaction of particles
with spherical nuclei in their elastic scattering (or in a
decay of compound spherical systems).

We point out a possibility of construction of space asymmetric
reflectionless potentials.
% ***************************************************************************

% ***************************************************************************
\section{Interdependences between spectral characteristics of
potentials-partners
\label{sec.2}}

Let's consider an one-dimensional case of movement of a particle
with mass $m$ in a potential field $V(x)$. We introduce the
operators $A$ and $A^{+}$ of the following form:
\begin{equation}
\begin{array}{ll}
  A =
  \displaystyle\frac{\hbar}{\sqrt{2m}}
  \displaystyle\frac{d}{dx}
  + W(x), &
  A^{+} =
  -\displaystyle\frac{\hbar}{\sqrt{2m}}
  \displaystyle\frac{d}{dx}
  + W(x),
\end{array}
\label{eq.2.1}
\end{equation}                                          %       (2.1)
where $W(x)$ is the function defined in the whole space region $x$.
Let's assume, that this function is continuous in the whole region
of its definition except for several possible points of
discontinuity.
On the basis of the operators $A$ and $A^{+}$ one can construct two
Hamiltonians for movement of this particle inside two different
potentials $V_{1}(x)$ and $V_{2}(x)$:
\begin{equation}
\begin{array}{l}
  H_{1} = A^{+} A =
  -\displaystyle\frac{\hbar^{2}}{2m}
  \displaystyle\frac{d^{2}}{dx^{2}}
  + V_{1}(x), \\
  H_{2} = A A^{+} =
  -\displaystyle\frac{\hbar^{2}}{2m}
  \displaystyle\frac{d^{2}}{dx^{2}}
  + V_{2}(x),
\end{array}
\label{eq.2.2}
\end{equation}                                          %       (2.2)
where the potentials $V_{1}(x)$ and $V_{2}(x)$ are determined as
follows:
\begin{equation}
\begin{array}{ll}
  V_{1}(x) =
  W^{2}(x) - \displaystyle\frac{\hbar}{\sqrt{2m}}
  \displaystyle\frac{d W(x)}{dx}, &
  V_{2}(x) =
  W^{2}(x) + \displaystyle\frac{\hbar}{\sqrt{2m}}
  \displaystyle\frac{d W(x)}{dx}.
\end{array}
\label{eq.2.3}
\end{equation}                                          %       (2.3)

In accordance with SUSY QM theory \cite{Cooper.1995.PRPLC}, the
function $W(x)$ is named as \emph{superpotential}, whereas the
potentials $V_{1}(x)$ and $V_{2}(x)$ are named as
\emph{supersymmetric potentials - partners} (or \emph{SUSY
potentials - partners}). A construction of the Hamiltonians of two
quantum systems on the basis of the same operators $A$ and $A^{+}$
interrelates dependence between spectral characteristics
(spectra of energy, wave functions) of these systems. One can see
a reason of such interdependence in that two potentials $V_{1}(x)$
and $V_{2}(x)$ are connected through one function $W(x)$:
\begin{equation}
  V_{2} (x) = V_{1}(x) +
    2\displaystyle\frac{\hbar}{\sqrt{2m}}
    \displaystyle\frac{d W(x)}{dx}.
\label{eq.2.4}
\end{equation}                                          %       (2.4)
% ***************************************************************************

% ***************************************************************************
\subsection{Systems with discrete and continuous energy spectra
\label{sec.2.1}}

If the energy spectra of two considered systems are discrete, then
one can write:
\begin{equation}
\begin{array}{l}
  H_{1} \varphi^{(1)}_{n} =
  A^{+} A \varphi^{(1)}_{n} =
  E^{(1)}_{n} \varphi^{(1)}_{n}, \\
  H_{2} \varphi^{(2)}_{n} =
  A A^{+} \varphi^{(2)}_{n} =
  E^{(2)}_{n} \varphi^{(2)}_{n},
\end{array}
\label{eq.2.1.1}
\end{equation}                                          %       (2.1.1)
where $E^{(1)}_{n}$ and $E^{(2)}_{n}$ are the energy levels with
number $n$ ($n$ is natural number) for two systems with potentials
$V_{1}(x)$ and $V_{2}(x)$,
$\varphi^{(1)}_{n}$ and $\varphi^{(2)}_{n}$ are the wave functions
(eigen-functions), corresponding to this levels.
From here we obtain:
\begin{equation}
\begin{array}{l}
  H_{2} (A \varphi^{(1)}_{n}) =
  A A^{+} A \varphi^{(1)}_{n} =
  E^{(1)}_{n} (A \varphi^{(1)}_{n}), \\
  H_{1} (A^{+} \varphi^{(2)}_{n}) =
  A^{+} A A^{+} \varphi^{(2)}_{n} =
  E^{(2)}_{n} (A^{+} \varphi^{(2)}_{n}).
\end{array}
\label{eq.2.1.2}
\end{equation}                                          %       (2.1.2)
Let's displace the potential $V_{1}(x)$ by such a way that
$E^{1}_{0}=0$ (it does not influence on a relative arrangement of
the levels in the energy spectra and shapes of the wave functions).
Analyzing (\ref{eq.2.1.2}), one can obtain the following
interdependences between the energy spectra and the wave functions
(see~\cite{Cooper.1995.PRPLC}, p.~275--276):
\begin{equation}
\begin{array}{ll}
  E^{(1)}_{0} = 0, &
  \varphi^{(2)}_{n} = (E^{(1)}_{n+1})^{-1/2} A \varphi^{(1)}_{n+1}, \\
  E^{(2)}_{n} = E^{(1)}_{n+1}, &
  \varphi^{(1)}_{n+1} = (E^{(2)}_{n})^{-1/2} A^{+} \varphi^{(2)}_{n}
\end{array}
\label{eq.2.1.3}
\end{equation}                                          %       (2.1.3)
(though other variants of the interdependences are possible also).
Here, a normalization condition for the wave functions is taken
into account for the discrete spectrum:
\begin{equation}
\begin{array}{ll}
  \displaystyle\int \bigl|\varphi^{(1)}_{n}(x) \bigr|^{2} \, dx = 1, &
  \displaystyle\int |\varphi^{(2)}_{n}(x)|^{2} \, dx = 1.
\end{array}
\label{eq.2.1.4}
\end{equation}                                          %       (2.1.4)

If the energy spectra of two systems are continuous, then one can
find the interdependences between their wave functions also. In
this case one can rewrite the system (\ref{eq.2.1.1}) by such a
way:
\begin{equation}
\begin{array}{l}
  H_{1} \varphi^{(1)}_{k} =
  A^{+} A \varphi^{(1)}_{k} =
  E^{(1)}_{k} \varphi^{(1)}_{k}, \\
  H_{2} \varphi^{(2)}_{k^{\prime}} =
  A A^{+} \varphi^{(2)}_{k^{\prime}} =
  E^{(2)}_{k^{\prime}} \varphi^{(2)}_{k^{\prime}},
\end{array}
\label{eq.2.1.5}
\end{equation}                                          %       (2.1.5)
where $E^{(1)}_{k}$ and $E^{(2)}_{k^{\prime}}$ are the energy
levels for two systems with the potentials $V_{1}(x)$ and $V_{2}(x)$,
having the continuous spectra of values,
$\varphi^{(1)}_{k}(x)$ and $\varphi^{(2)}_{k^{\prime}}(x)$,
$k = \displaystyle\frac{1}{\hbar}\sqrt{2mE^{(1)}_{k}}$ and
$k^{\prime} =
\displaystyle\frac{1}{\hbar}\sqrt{2mE^{(2)}_{k^{\prime}}}$ are
the wave functions and the wave vectors, corresponding to the
levels $E^{(1)}_{k}$ and $E^{(2)}_{k^{\prime}}$.
From (\ref{eq.2.1.5}) we obtain:
\begin{equation}
  H_{2} (A \varphi^{(1)}_{k}) =
  A A^{+} (A \varphi^{(1)}_{k}) =
  A (A^{+} A \varphi^{(1)}_{k}) =
  A (E^{(1)}_{k} \varphi^{(1)}_{k}) =
  E^{(1)}_{k} (A \varphi^{(1)}_{k}).
\label{eq.2.1.6}
\end{equation}                                          %       (2.1.6)
And one can write:
\begin{equation}
\begin{array}{lcr}
  A \varphi^{(1)}_{k} (x) =
    C_{2} \varphi^{(2)}_{k^{\prime}} (x), &
  E^{(1)}_{k} =
    \displaystyle\frac{E^{(2)}_{k^{\prime}}}{C_{2}}, &
  C_{2} = const.
\end{array}
\label{eq.2.1.7}
\end{equation}                                          %       (2.1.7)
Considering the second expression of this system and taking into
account the definitions of the wave vectors $k$ and $k^{\prime}$,
one can obtain the constant $C_{2}$.
%
% \begin{equation}
%   C_{2} =
%   \displaystyle\frac{E^{(2)}_{k^{\prime}}}{E^{(1)}_{k}} =
%   \displaystyle\frac{{k^{\prime}}^{2}}{k^{2}}.
% \label{eq.2.1.8}
% \end{equation}                                          %       (2.1.8)
% 
Taking into account (\ref{eq.2.1.5}), we write:
\begin{equation}
  H_{1} (A^{+} \varphi^{(2)}_{k^{\prime}}) =
  A^{+} A A^{+} \varphi^{(2)}_{k^{\prime}} =
%  A^{+} (A A^{+} \varphi^{(2)}_{k^{\prime}}) =
  A^{+} (E^{(2)}_{k^{\prime}} \varphi^{(2)}_{k^{\prime}}) =
  E^{(2)}_{k^{\prime}} (A^{+} \varphi^{(2)}_{k^{\prime}}),
\label{eq.2.1.9}
\end{equation}                                          %       (2.1.9)
and find:
\begin{equation}
\begin{array}{lcr}
  A^{+} \varphi^{(2)}_{k^{\prime}} (x) =
    C_{1} \varphi^{(1)}_{k} (x), &
  E^{(2)}_{k^{\prime}} =
    \displaystyle\frac{E^{(1)}_{k}}{C_{1}}, &
  C_{1} =
    \displaystyle\frac{1}{C_{2}} =
    \displaystyle\frac{E^{(1)}_{k}}{E^{(2)}_{k^{\prime}}} =
    \displaystyle\frac{{k}^{2}}{{k^{\prime}}^{2}}.
\end{array}
\label{eq.2.1.10}
\end{equation}                                          %       (2.1.10)

So, we have received the following interdependences between the
wave functions and the energy levels for SUSY systems-partners in
the continuous energy spectra:
\begin{equation}
\begin{array}{lcr}
  \varphi^{(1)}_{k} (x) =
    \displaystyle\frac{{k^{\prime}}^{2}}{{k}^{2}}
    A^{+} \varphi^{(2)}_{k^{\prime}} (x), &
  \varphi^{(2)}_{k^{\prime}} (x) =
    \displaystyle\frac{{k}^{2}}{{k^{\prime}}^{2}}
    A \varphi^{(1)}_{k} (x), &
  E^{(1)}_{k} =
    \displaystyle\frac{{k}^{2}}{{k^{\prime}}^{2}}
    E^{(2)}_{k^{\prime}}.
\end{array}
\label{eq.2.1.11}
\end{equation}                                          %       (2.1.11)
One can see that lack of coincidence of the levels of the
continuous energy spectra of SUSY systems-partners is possible
(unlike the SUSY systems-partners with the discrete spectra).
For determination of the exact interdependence between coefficients
$k$ and $k^{\prime}$ it is necessary to use the normalization
condition of the wave functions (for the continuous energy spectra)
with taking into account of boundary conditions.
% ***************************************************************************

% ***************************************************************************
\subsection{Interdependences between coefficients of penetrability
and reflection
\label{sec.2.2}}

The SUSY QM methods allow to find the interdependences between
coefficients of penetrability and reflection for two SUSY
systems-partners in the regions of the continuous energy spectra
(for example, see~\cite{Cooper.1995.PRPLC}, p.~278--279).
Let the superpotential $W(x)$ and the potentials $V_{1}(x)$ and
$V_{2}(x)$ converge at $x \to \pm\infty$ to finite limits:
\begin{equation}
\begin{array}{ll}
  W (x) \to W_{\pm}, &
  V_{1} (x) = V_{2} (x) \to W^{2}_{\pm}.
\end{array}
\label{eq.2.2.2}
\end{equation}                                          %       (2.2.2)
Let's consider a propagation of a plane wave $e^{ik_{-}x}$ in the
positive direction along the axis $x$ in the fields of the
potentials $V_{1}(x)$ and $V_{2}(x)$ (we assume, that the
propagation inside these two potentials occurs along identical
levels $E^{(1)}$ and $E^{(2)}$). The incident wave from the left
gives transmitted waves $T_{1}(k_{-}, k_{+}) e^{ik_{+}x}$ and
$T_{2}(k_{-}, k_{+}) e^{ik_{+}x}$, and also reflected waves
$R_{1}(k_{-}) e^{-ik_{-}x}$ and $R_{2}(k_{-}) e^{-ik_{-}x}$.
We have:
\begin{equation}
\begin{array}{ll}
  \varphi^{(1)}(k, x \to -\infty) \to
    N_{1} (e^{ik_{-}x} + R_{1} e^{-ik_{-}x}), &
  \varphi^{(1)}(k, x \to +\infty) \to
    N_{1} T_{1} e^{ik_{+}x}, \\
  \varphi^{(2)}(k, x \to -\infty) \to
    N_{2} (e^{ik_{-}x} + R_{2} e^{-ik_{-}x}), &
  \varphi^{(2)}(k, x \to +\infty) \to
    N_{2} T_{2} e^{ik_{+}x},
\end{array}
\label{eq.2.2.3}
\end{equation}                                          %       (2.2.3)
where $k$, $k_{-}$ and $k_{+}$ are determined by such a way:
\begin{equation}
\begin{array}{lcr}
  k = \displaystyle\frac{1}{\hbar} \sqrt{2mE}, &
  k_{-} = \displaystyle\frac{1}{\hbar} \sqrt{2m(E-W^{2}_{-})}, &
  k_{+} = \displaystyle\frac{1}{\hbar} \sqrt{2m(E-W^{2}_{+})},
\end{array}
\label{eq.2.2.4}
\end{equation}                                          %       (2.2.4)
and the coefficients $N_{1}$ and $N_{2}$ can be found from the
normalization conditions with taking into account of the potentials
forms and the boundary conditions.

Using the interdependences (\ref{eq.2.1.11}) between the wave
functions of two systems with the continuous spectra, we write:
\begin{equation}
\begin{array}{l}
  e^{ik_{-}x} + R_{1} e^{-ik_{-}x} =
    C_{2} N_{2} \biggl[
    \biggl(-\displaystyle\frac{ik_{-}\hbar}{\sqrt{2m}} + W_{-}\biggr)
    e^{ik_{-}x} +
    \biggl(\displaystyle\frac{ik_{-}\hbar}{\sqrt{2m}} + W_{-}\biggr)
    e^{-ik_{-}x}
    R_{2}\biggr], \\
  T_{1} e^{ik_{+}x} =
    C_{2} N_{2}
    \biggl(-\displaystyle\frac{ik_{+}\hbar}{\sqrt{2m}} + W_{+}\biggr)
     e^{ik_{+}x} T_{2},
\end{array}
\label{eq.2.2.5}
\end{equation}                                          %       (2.2.5)
where $C_{2}$ is a constant determined in (\ref{eq.2.1.10}).
Equating items in (\ref{eq.2.2.5}) at the same exponents, one can
find:
%
% \begin{equation}
%   C_{2} N_{2} =
%     \displaystyle\frac
%     {1}{W_{-} - \displaystyle\frac{ik_{-}\hbar}{\sqrt{2m}}}
% \label{eq.2.2.6}
% \end{equation}                                          %       (2.2.6)
%
\begin{equation}
\begin{array}{ll}
  C_{2} N_{2} =
    \displaystyle\frac
    {1}{W_{-} - \displaystyle\frac{ik_{-}\hbar}{\sqrt{2m}}}, & \\
  R_{1}(k_{-}) =
    R_{2}(k_{-}) \displaystyle\frac
    {W_{-} + \displaystyle\frac{ik_{-}\hbar}{\sqrt{2m}}}
    {W_{-} - \displaystyle\frac{ik_{-}\hbar}{\sqrt{2m}}}, &
  T_{1}(k_{-}, k_{+}) =
    T_{2}(k_{-}, k_{+}) \displaystyle\frac
    {W_{+} - \displaystyle\frac{ik_{+}\hbar}{\sqrt{2m}}}
    {W_{-} - \displaystyle\frac{ik_{-}\hbar}{\sqrt{2m}}}.
\end{array}
\label{eq.2.2.7}
\end{equation}                                          %       (2.2.7)
The expressions (\ref{eq.2.2.7}) show the interdependences between
amplitudes of the transmission and the reflection for SUSY
systems-partners. Coefficients of the penetrability and the
reflection for the potentials $V_{1}(x)$ and $V_{2}(x)$ can be
calculated as squares of modules of the amplitudes of the
transmission and the reflection.
% ***************************************************************************

% ***************************************************************************
\subsection{Isospectral potentials
\label{sec.2.3}}

If we know the potential $V_{1}(x)$ only, then we can find the
superpotential $W(x)$ by solving the first equation of the system
(\ref{eq.2.3}). It is the differential equation of the first order.
One can calculate the function $W(x)$ with an accuracy to one
arbitrary constant of integration, which can be considered as a
free parameter. Therefore, the function $W(x)$ is
\emph{ambiguously determined}. Generally, changing this free
parameter, one can change a form of the function $W(x)$ without
displacement of the levels in the energy spectrum and without
change of a form of the potential $V_{1}(x)$.

According to (\ref{eq.2.4}), the potential $V_{2}(x)$ is expressed
through $V_{1}(x)$ and derivative of $W(x)$ on coordinate. As the
first equation of the system (\ref{eq.2.3}) is nonlinear, generally
we can not speak that change of the considered above free parameter,
resulting in the change of the superpotential $W(x)$, does not
change the potential $V_{2}(x)$. On the other side, one can change
the form of the potential - partner $V_{2}(x)$ by changing this
parameter. Therefore, one can construct a set of potentials which
have identical spectra of energy, different shapes and are named as
\emph{isospectral potentials}.
As a confirmation of existence of the isospectral potentials,
there is a beautiful example of one-parameter family of the
one-dimensional isospectral potentials constructed on the basis of
the oscillator potential and having identical equidistant spectrum
of energy (see~\cite{Cooper.1995.PRPLC}, p.~326).

Let's assume that the potential $V_{1}(x)$ is known and defined
uniquely. Then we make the following conclusions:
\begin{itemize}

\item
All possible types of the potentials $V_{2}(x)$, which are
SUSY-partners to the potential $V_{1}(x)$, make one-parameter set
$\mathcal{A}$ of the isospectral potentials (connected through
one free parameter).

\item
Any potential $V_{2}(x)$, which is the SUSY-partner to the
potential $V_{1}(x)$, belongs to this one-parameter set
$\mathcal{A}$ of the isospectral potentials.

\item
One can suppose that any potential $V_{2}(x)$, which belongs to
one-parameter set $\mathcal{A}$ of the isospectral potentials, is
the SUSY-partner to the potential $V_{1}(x)$.
\end{itemize}
\emph{If the third item is carried out, then we come to analogy
(one-to-one relationship) between one-parameter set of the
isospectral potentials and the set of all possible types of the
potentials, which are the SUSY-partners to one uniquely chosen
potential}. This analogy is obtained for the first time (though
some dependences between these two sets were studied earlier).
On its basis one can propose a choice of the parameter, concerning
which the set of the one-parameter isospectral potentials is
determined.

These reasoning are fulfilled both for systems with the discrete
spectra of energy, and for systems with the continuous spectra of
energy. They can be continued if to consider SUSY-hierarchy, to
which the potentials $V_{1}(x)$ and $V_{2}(x)$ belong. So,
finding possible types of the potentials - partners to the
potential $V_{2}(x)$, we obtain a set new isospectral potentials
$V_{3}(x)$, which are determined relatively the potential
$V_{2}(x)$ and connected among themselves through one free
parameter or determined relatively the initial potential $V_{1}(x)$
and connected among themselves through two independent free
parameters.

Thus, we come to the conclusion about that \emph{some isospectral
potentials are connected through only one independent free parameter,
others isospectral potentials --- through two independent free
parameters, third --- through three and so on}
(another approach to a construction of one- and many-parameter
families of the isospectral potentials, where an analysis of an
obvious form of wave functions is used, can be found in
\cite{Cooper.1995.PRPLC}, p.~323--329, with references to other
papers).
% ***************************************************************************

% ***************************************************************************
\section{Search of a general form of the reflectionless potential
\label{sec.3}}

According to (\ref{eq.2.2.7}), if we know a reflectionless
potential (its reflection coefficient is equal to zero), than its
SUSY potentials-partners can be also reflectionless.

{\bf Rule }:
\emph{
two potentials $V_{1}(x)$ and $V_{2}(x)$ are reflectionless only
when:}
\begin{itemize}

\item
\emph{these potentials are SUSY potentials-partners;}

\item
\emph{asymptotic expressions of wave functions for both potentials
have the form (\ref{eq.2.2.3}) (it is pointed out for the first time).}
\end{itemize}

Using this natural rule and knowing a form of one reflectionless
potential only, one can construct a set of new reflectionless
potentials. Let's apply such an approach to construction of the new
reflectionless potentials, having taken as first the constant
potential of the form:
\begin{equation}
  V(x) = A^{2} = \mbox{const}
\label{eq.3.1}
\end{equation}                                          %       (3.1)
(we study the cases $A^{2} \le 0$ and $A^{2} \ge 0$ at
$A^{2} \in Re$). A plane wave $e^{ikx}$ (where $k$ is a wave vector
determined at $E-V$) in the field of such potential does not meet
an obstacle during its propagation. One can think that the
potential (\ref{eq.3.1}) is reflectionless for the propagation of
this wave (one can make sure in this by calculating the coefficients
of the penetrability and the reflection).

If to construct SUSY-hierarchy, which has the potential
(\ref{eq.3.1}), than only those potentials of such hierarchy are
reflectionless, the asymptotic forms of the wave functions for
which look like (\ref{eq.2.2.3}).
All such potentials make one SUSY-hierarchy of the reflectionless
potentials. Here, \emph{a reason of the absolute transparency of
any such potential can be seen in its SUSY-interrelation with the
constant potential}.
% ***************************************************************************

% ***************************************************************************
\section{Potentials-partners to the constant potential
\label{sec.4}}

Let's assume, that the SUSY-hierarchy for the potential
(\ref{eq.3.1}) is constructed. Let this potential be $V_{n}(x)$
with number $n$ in this hierarchy. We shall find the nearest
potentials to it in this hierarchy, i.~e. its potentials - partners.

% ***************************************************************************
\subsection{The potentials-partners with a number ``up''
\label{sec.4.1}}

At first we shall find the potential - partner $V_{n+1}(x)$ with
the number ``up'', i.~e. when interdependence between these
potentials has a form:
\begin{equation}
\begin{array}{ll}
  V_{n}(x) =
  W_{n}^{2}(x) - \displaystyle\frac{\hbar}{\sqrt{2m}}
  \displaystyle\frac{d W_{n}(x)}{dx}, &
  V_{n+1}(x) =
  W_{n}^{2}(x) + \displaystyle\frac{\hbar}{\sqrt{2m}}
  \displaystyle\frac{d W_{n}(x)}{dx}.
\end{array}
\label{eq.4.1.1}
\end{equation}                                          %       (4.1.1)

Taking into account (\ref{eq.3.1}), we write the differential
equation for calculating superpotential $W_{n}(x)$:
\begin{equation}
  W_{n}^{2}(x) - \displaystyle\frac{\hbar}{\sqrt{2m}}
  \displaystyle\frac{d W_{n}(x)}{dx} = A^{2}.
\label{eq.4.1.2}
\end{equation}                                          %       (4.1.2)
Introduce designation:
\begin{equation}
  \alpha = \displaystyle\frac{\hbar}{\sqrt{2m}},
\label{eq.4.1.3}
\end{equation}                                          %       (4.1.3)
and solve the equation (\ref{eq.4.1.2}). At $W_{n}^{2}(x) = A^{2}$
we obtain:
%
% \begin{equation}
%   \left\{
%   \begin{array}{ll}
%      W_{n}^{2}(x) - A^{2} = 0; & \\
%      dx = \alpha \displaystyle\frac{d W_{n}(x)}{W_{n}^{2}(x)-A^{2}}, &
%      \mbox{при } W_{n}^{2}(x) - A^{2} \ne 0;
%   \end{array}
%   \right.
% \label{eq.4.1.4}
% \end{equation}                                          %       (4.1.4)
%
\begin{equation}
  V_{n+1}(x) = W_{n}^{2}(x) = V_{n}(x) = A^{2}.
\label{eq.4.1.5}
\end{equation}                                          %       (4.1.5)
We see that in this case the potential $V_{n}(x)$ is supersymmetric
to itself.

At $W_{n}^{2}(x) \ne A^{2}$ we obtain:
\begin{equation}
% \begin{array}{lll}
%   \biggl( dx = \alpha \displaystyle\frac{d W_{n}(x)}
%                 {W_{n}^{2}(x)-A^{2}} \biggr)
%   & \Longrightarrow &
  \displaystyle\int dx =
  \alpha \displaystyle\int \displaystyle\frac{d W_{n}(x)}{W_{n}^{2}(x)-A^{2}}.
% \end{array}
\label{eq.4.1.6}
\end{equation}                                          %       (4.1.6)
We find a dependence of the function $W_{n}$ on $x$. Therefore, we
have the indefinite integral, with taking into account a constant of
integration. At integration of this equation we obtain three
different cases.

1) The case $A=0$. Then:
\begin{equation}
  W_{n}(x) =
    - \displaystyle\frac{\alpha}{x + x_{0}},
    \mbox{ and } x+x_{0} \ne 0.
\label{eq.4.1.7}
\end{equation}                                          %       (4.1.7)
Here, the constant of integration $x_{0}$ is introduced.

2) The case $A \ne 0$ and $A^{2} > 0$. Then:
\begin{equation}
  A^{2} = (\rho e^{i\phi})^{2} = \rho^{2} e^{2i\phi} = +\rho^{2} =
  |A|^{2},
\label{eq.4.1.8}
\end{equation}                                          %       (4.1.8)
\begin{equation}
  \left\{
  \begin{array}{ll}
    \displaystyle\frac{W_{n}-|A|}{W_{n}+|A|} =
    + e^{\displaystyle\frac{2|A|}{\alpha}(x+x_{0})}, &
    \mbox{ at } |W_{n}| > |A|; \\
    \displaystyle\frac{W_{n}-|A|}{W_{n}+|A|} =
    - e^{\displaystyle\frac{2|A|}{\alpha}(x+x_{0})}, &
    \mbox{ at } |W_{n}| < |A|.
  \end{array}
  \right.
\label{eq.4.1.9}
\end{equation}                                          %       (4.1.9)
Let's consider the first equation of this system. At $x+x_{0} \ne 0$
its solution has a form:
\begin{equation}
\begin{array}{l}
  W_{n}(x) =
%    -|A| \displaystyle\frac
%    {e^{\displaystyle\frac{|A|}{\alpha}(x+x_{0})} +
%    e^{-\displaystyle\frac{|A|}{\alpha}(x+x_{0})}}
%    {e^{\displaystyle\frac{|A|}{\alpha}(x+x_{0})} -
%    e^{-\displaystyle\frac{|A|}{\alpha}(x+x_{0})}} = 
    -|A| \coth\biggl({\displaystyle\frac{|A|}{\alpha}(x+x_{0})}\biggr).
\end{array}
\label{eq.4.1.12}
\end{equation}                                          %       (4.1.12)
At $x+x_{0} = 0$ two cases are possible:
\[
\begin{array}{ccl}
  |A| = 0 & \Longrightarrow & \mbox{ false}; \\
  W_{n}(x \to -x_{0}) \to \pm\infty & \Longrightarrow & \mbox{ true}.
\end{array}
\label{eq.4.1.13}
\]                                                      %       (4.1.13)
The first condition is not carried out, because of we study the
case $A \ne 0$. The second condition is carried out automatically
as a limit of the expression (\ref{eq.4.1.12}) and satisfies to the
condition: $|W_{n}(x)| > |A|$. Therefore, one can remove the
restriction $x+x_{0} \ne 0$ on the solution (\ref{eq.4.1.12}).

The second equation of the system (\ref{eq.4.1.9}) gives such result:
\begin{equation}
    W_{n} (x) =
%    |A| \displaystyle\frac
%    {1 - e^{\displaystyle\frac{2|A|}{\alpha}(x+x_{0})}}
%    {1 + e^{\displaystyle\frac{2|A|}{\alpha}(x+x_{0})}} =
    -|A| \tanh 
    \biggl(\displaystyle\frac{|A|}{\alpha}(x+x_{0})\biggr)
\label{eq.4.1.14}
\end{equation}                                          %       (4.1.14)
and does not use the restriction $x \ne -x_{0}$ also (here, $W_{n}(x)
\to 0$ at $x+x_{0} \to 0$).

3) The case $A \ne 0$ and $A^{2} < 0$. Then:
\begin{equation}
  A^{2} = (\rho e^{i\phi})^{2} = \rho^{2} e^{2i\phi} = -\rho^{2} =
  -|A|^{2},
\label{eq.4.1.15}
\end{equation}                                          %       (4.1.15)
\begin{equation}
  W_{n}(x) =
    |A| \tan\biggl(\displaystyle\frac{|A|}{\alpha}(x + x_{0})\biggr).
\label{eq.4.1.16}
\end{equation}                                          %       (4.1.16)

So, we have obtained five solutions of the function $W_{n}(x)$:
\begin{equation}
% \begin{array}{l}
  W_{n}(x) =
  \left\{
  \begin{array}{cl}
% 1
    \pm |A| = const; & \\
% 2
    - \displaystyle\frac{\alpha}{x + x_{0}},
    & \mbox{for } A=0, W_{n}^{2}(x) \ne 0 \mbox{ and }\\
    & x+x_{0} \ne 0; \\
% 3
    -|A| \coth 
      \biggl(\displaystyle\frac{|A|}{\alpha}(x+x_{0})\biggr),
    & \mbox{for } A^{2} > 0 \mbox{ and } |W_{n}(x)| > |A|; \\
% 4
    -|A| \tanh
      \biggl(\displaystyle\frac{|A|}{\alpha}(x+x_{0})\biggr),
    & \mbox{for } A^{2} > 0 \mbox{ and } |W_{n}(x)| < |A|; \\
% 5
    |A| \tan \biggl(
      \displaystyle\frac{|A|}{\alpha}(x + x_{0}) \biggl),
    & \mbox{for } A^{2} < 0 \mbox{ and } W_{n}^{2}(x) \ne A^{2},
  \end{array}
  \right.
% \end{array}
\label{eq.4.1.17}
\end{equation}                                          %       (4.1.17)
where $x_{0}$ is the constant of integration (unique free
parameter).

According to (\ref{eq.4.1.1}), the superpotential $W(x)$ uniquely
defines the potential $V_{n+1}(x)$, which has five solutions:
\begin{equation}
\begin{array}{l}
  V_{n+1}(x) =
  \left\{
  \begin{array}{cl}
% 1
    A^{2}, & \mbox{for } W^{2}_{n}(x) = A^{2}; \\
% 2
    \displaystyle\frac{2\alpha^{2}}{(x+x_{0})^{2}},
    & \mbox{for } A=0, W_{n}^{2}(x) \ne 0 \mbox{ and } \\
    & x+x_{0} \ne 0; \\
% 3
    |A|^{2} \left(1 + \displaystyle\frac{2}{\sinh^{2}
      \bigl(|A|(x+x_{0}) / \alpha \bigr)} \right),
    & \mbox{for } A^{2} > 0 \mbox{ and } |W_{n}(x)| > |A|; \\
% 4
    |A|^{2} \left(1 - \displaystyle\frac{2}{\cosh^{2}
      \bigl(|A|(x+x_{0}) / \alpha \bigr)} \right),
    & \mbox{for } A^{2} > 0 \mbox{ and } |W_{n}(x)| < |A|; \\
% 5
    |A|^{2} \left(\displaystyle\frac{2}{\cos^{2}
      \bigl(|A|(x+x_{0}) / \alpha \bigr)} - 1 \right),
    & \mbox{for } A^{2} < 0 \mbox{ and } W_{n}^{2}(x) \ne A^{2}.
  \end{array}
  \right.
\end{array}
\label{eq.4.1.22}
\end{equation}                                          %       (4.1.22)
% ***************************************************************************

% ***************************************************************************
\subsection{The potentials-partners with a number ``down''
\label{sec.4.2}}

Now we find a potential $V_{n-1}(x)$ with the number ``down'',
which is the SUSY-partner to the constant potential $V_{n}(x)$ of
the form (\ref{eq.3.1}). An equation for determination of a
superpotential $W_{n-1}(x)$, connecting $V_{n-1}(x)$ and
$V_{n}(x)$, has a form:
\begin{equation}
  W^{2}_{n-1}(x) + \alpha \displaystyle\frac{d W_{n-1}(x)}{dx} = A^{2}.
\label{eq.4.2.2}
\end{equation}                                          %       (4.2.2)
At replacement
\begin{equation}
  \bar{\alpha} \to -\alpha.
\label{eq.4.2.3}
\end{equation}                                          %       (4.2.3)
this equation transforms into the equation (\ref{eq.4.1.2}). Then
the solutions of the equation (\ref{eq.4.1.17}) become the solutions
of the equation (\ref{eq.4.2.2}):
\begin{equation}
  W_{n-1}(x) =
  \left\{
  \begin{array}{cl}
% 1
    \pm |A| = const; & \\
% 2
    \displaystyle\frac{\alpha}{x + x_{0}},
    & \mbox{at } A=0, W_{n-1}^{2}(x) \ne 0 \mbox{ and } \\
    & x+x_{0} \ne 0; \\
% 3
    |A| \coth
      \biggl(\displaystyle\frac{|A|}{\alpha}(x+x_{0})\biggr),
    & \mbox{at } A^{2} > 0 \mbox{ and } |W_{n-1}(x)| > |A|; \\
% 4
    |A| \tanh
      \biggl(\displaystyle\frac{|A|}{\alpha}(x+x_{0})\biggr),
    & \mbox{at } A^{2} > 0 \mbox{ and } |W_{n-1}(x)| < |A|; \\
% 5
    -|A| \tan \biggl(
      \displaystyle\frac{|A|}{\alpha}(x + x_{0}) \biggl),
    & \mbox{at } A^{2} < 0 \mbox{ and } W_{n-1}^{2}(x) \ne A^{2},
  \end{array}
  \right.
% \end{array}
\label{eq.4.2.4}
\end{equation}                                          %       (4.2.4)
where $x_{0}$ is the constant of integration (unique free
parameter), and the solutions for $V_{n-1}(x)$ coincide with the
solutions for $V_{n+1}(x)$:
\begin{equation}
  V_{n-1}(x) = V_{n+1}(x).
\label{eq.4.2.5}
\end{equation}                                          %       (4.2.5)

So, we have obtained the SUSY potentials-partners to the constant
potential (\ref{eq.3.1}). Note the following:
\begin{itemize}
\item
All possible types of the SUSY potentials-partners with the numbers
``up'' and ``down'' to the constant potential are defined by the
expressions (\ref{eq.4.1.22}) and (\ref{eq.4.2.5}). There are no
any other type (with the exception of these five solutions) of the
potential-partner to the constant potential
(it is obtained for the first time).

\item
In order to find out, which potentials from the obtained above
solutions are reflectionless, it is need to analyze an asymptotic
form of wave functions of these potentials.
\begin{itemize}
\item
The first potential is constant. It is supersymmetric to itself,
keeps a property of absolute transparency and does not give
anything new.

\item
The second potential has the inverse power dependence on coordinate.
One can construct new reflectionless potentials on the basis of it
(where there is a tunneling) and it will be studied in
subsection~\ref{sec.5.1}.

\item
The fourth potential is known in the literature as reflectionless
potential (see~p.~280 in \cite{Cooper.1995.PRPLC}). It is used often
also in construction and analysis of one- and many-soliton
reflectionless potentials (see~p.~328 in \cite{Cooper.1995.PRPLC}).

\item
If the fifth potential is determined on the whole axis $x$, then
in a general case it has an infinite set of barriers (like
$\delta$-barriers), and it is not clear, how to consider a directed
motion of a particle (with tunneling) in the field of such potential.
At limits $x \to \pm\infty$ this potential does not converges to
unambiguous values. Therefore, its wave function does not have the
unambiguous form in the asymptotic areas and is not reduced to the
form (\ref{eq.2.2.3}) in a general case.
Examples of this potential as potential-partner to the constant
potential defined on the given finite region on $x$ is studied in
literature (see~p.~278--279 in \cite{Cooper.1995.PRPLC}).
\end{itemize}
The second and third potentials as reflectionless are not found by us
in other papers.

\item
At $x \to \pm\infty$ the wave functions tend to the form
(\ref{eq.2.2.3}) only for those potentials, which have finite
values in a whole region of their definition and converge to
unique finite values at $x \to \pm\infty$.
Therefore, for extraction of the reflectionless potentials from
the considered above SUSY-hierarchy the analysis of the form of
the wave function in the asymptotic areas can be replaced by the
analysis of existence of the potential finiteness in the whole
region of its definition and existence of its unique finite limits
at $x \to \pm\infty$.

\item
One can see $x_{0}$ as a free parameter, changing which one can
change the forms of the potentials (with displacement along the
$x$ axis) without displacement of the levels in the energy spectra
(such potentials make one-parameter set of isospectral potentials).
Here, if the absolute transparency exists, then it is kept (in the
regions of application of these potentials).
The same situation exists in the one-parameter family of the
isospectral potentials constructed on the basis of the analysis of
wave functions (for example, see~p.~326 and Fig.~7.1
in~\cite{Cooper.1995.PRPLC}).

\item
All found reflectionless potentials have a simple analytical form
and are expressed through elementary functions.
\end{itemize}
% ***************************************************************************

% ***************************************************************************
\section{Construction of the SUSY-hierarchy of the reflectionless
potentials
\label{sec.5}}

Consistently determining by a recurrent way the potentials
belonging to one SUSY-hierarchy, which contains constant potential,
and selecting the potentials of finite height with unique finite
asymptotic limits from this hierarchy, one can find new types of
the reflectionless potentials. Let's consider some cases of
the reflectionless SUSY-hierarchies.
% ***************************************************************************

% ***************************************************************************
\subsection{Inverse power reflectionless potentials
\label{sec.5.1}}

Let's analyze a case, when a potential having the inverse power
dependence on spatial coordinate, can be reflectionless.
% ***************************************************************************

% ***************************************************************************
\subsubsection{Potentials-partners to the constant potential
\label{sec.5.1.1}}

Let's consider a superpotential of the form:
\begin{equation}
W(x) = \left\{
\begin{array}{ll}
   \displaystyle\frac{-\alpha}{x-x_{0}}, & \mbox{at } x<0; \\
   \displaystyle\frac{-\alpha}{x+x_{0}}, & \mbox{at } x>0;
\end{array} \right.
\label{eq.5.1.1.1}
\end{equation}                                          %       (5.1.1.1)
where $\alpha > 0$, $x_{0} > 0$. According to (\ref{eq.2.3}), the
function $W(x)$ uniquely defines the potentials-partners $V_{1}(x)$
and $V_{2}(x)$:
\begin{equation}
\mbox{for } x < 0 \left\{
\begin{array}{l}
   V_{1}(x) =
        W^{2}(x) - \displaystyle\frac{\hbar}{\sqrt{2m}}
        \displaystyle\frac{d W(x)}{dx} =
%         \displaystyle\frac{\alpha^{2}}{(x+x_{0})^{2}} -
%         \displaystyle\frac{\hbar}{\sqrt{2m}}
%         \displaystyle\frac{\alpha}{(x+x_{0})^{2}} =
        \displaystyle\frac{\alpha}{(x-x_{0})^{2}}
        \biggl(\alpha - \displaystyle\frac{\hbar}{\sqrt{2m}}\biggl), \\
   V_{2}(x) =
        W^{2}(x) + \displaystyle\frac{\hbar}{\sqrt{2m}}
        \displaystyle\frac{d W(x)}{dx} =
%         \displaystyle\frac{\alpha^{2}}{(x+x_{0})^{2}} +
%         \displaystyle\frac{\hbar}{\sqrt{2m}}
%         \displaystyle\frac{\alpha}{(x+x_{0})^{2}} =
        \displaystyle\frac{\alpha}{(x-x_{0})^{2}}
        \biggl(\alpha + \displaystyle\frac{\hbar}{\sqrt{2m}}\biggl);
\end{array} \right.
\label{eq.5.1.1.2}
\end{equation}                                          %       (5.1.1.2)
\begin{equation}
\mbox{for } x > 0 \left\{
\begin{array}{l}
   V_{1}(x) =
        W^{2}(x) - \displaystyle\frac{\hbar}{\sqrt{2m}}
        \displaystyle\frac{d W(x)}{dx} =
%        \displaystyle\frac{\alpha^{2}}{(x-x_{0})^{2}} -
%        \displaystyle\frac{\hbar}{\sqrt{2m}}
%        \displaystyle\frac{\alpha}{(x-x_{0})^{2}} =
        \displaystyle\frac{\alpha}{(x+x_{0})^{2}}
        \biggl(\alpha - \displaystyle\frac{\hbar}{\sqrt{2m}}\biggl), \\
   V_{2}(x) =
        W^{2}(x) + \displaystyle\frac{\hbar}{\sqrt{2m}}
        \displaystyle\frac{d W(x)}{dx} =
%        \displaystyle\frac{\alpha^{2}}{(x-x_{0})^{2}} +
%        \displaystyle\frac{\hbar}{\sqrt{2m}}
%        \displaystyle\frac{\alpha}{(x-x_{0})^{2}} =
        \displaystyle\frac{\alpha}{(x+x_{0})^{2}}
        \biggl(\alpha + \displaystyle\frac{\hbar}{\sqrt{2m}}\biggl).
\end{array} \right.
\label{eq.5.1.1.3}
\end{equation}                                          %       (5.1.1.3)

\begin{figure}[t]
\begin{center}
\includegraphics[width=5cm]{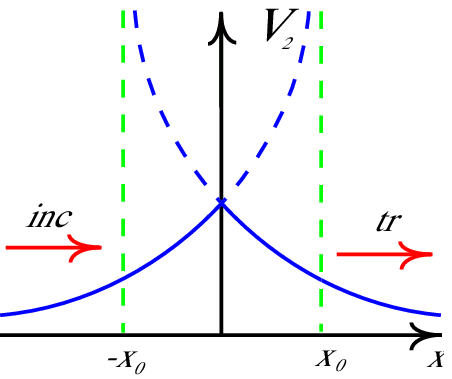}
\end{center}
\caption{The barrier becomes reflectionless at
  $\alpha = \displaystyle\frac{\hbar}{\sqrt{2m}}$
\label{fig.5111}}
\end{figure}

From (\ref{eq.5.1.1.2}) and (\ref{eq.5.1.1.3}) one can see, that
when the following condition
\begin{equation}
  \alpha = \displaystyle\frac{\hbar}{\sqrt{2m}}
\label{eq.5.1.1.4}
\end{equation}                                          %       (5.1.1.4)
is fulfilled, then the potential $V_{1}$ becomes zero. The
coefficient of penetrability of this potential for a propagation of
a plane wave is equal to one and, this sense, the potential
$V_{1}(x)$ is reflectionless.
The potential $V_{2}(x)$ is defined and is finited on the whole
axis $x$, has a barrier with a finite top at $x=x_{0}$ and tends
to zero in the asymptotic areas (see~Fig.~\ref{fig.5111}).
\emph{Therefore, an unidirectional movement of the plane wave with
tunneling in this potential is possible}.
According to (\ref{eq.2.2.7}), the coefficient of penetrability of
this potential also is equal to one:
\begin{equation}
  |T_{1}|^{2} = |T_{2}|^{2} = 1.
\label{eq.5.1.1.5}
\end{equation}                                          %       (5.1.1.5)

\begin{figure}[t]
\begin{center}
\includegraphics[width=5cm]{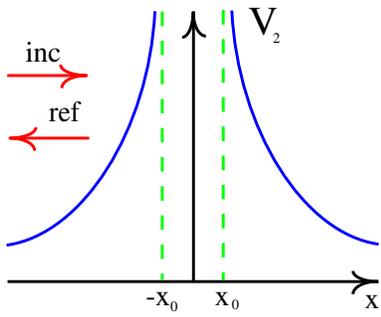}
\end{center}
\caption{Absolute reflection at $x_{0}<0$
\label{fig.5112}}
\end{figure}

\begin{figure}[t]
\begin{center}
\includegraphics[width=5cm]{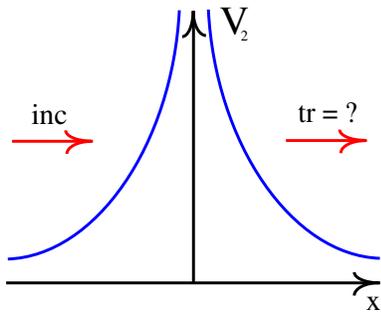}
\end{center}
\caption{Boundary case at $x_{0}=0$
\label{fig.5113}}
\end{figure}

Note a property: the coefficient of penetrability for the
reflectionless potential $V_{2}(x)$ does not changed with
displacement of $x_{0}$ (at $x_{0} > 0$).
At $x_{0} < 0$ there is a region $x \in ]-|x_{0}|, +|x_{0}|[$,
where this potential becomes indefinitely large and absolutely
opaque (see~Fig.~\ref{fig.5112}).
A case $x_{0}=0$ is boundary (see~Fig.~\ref {fig.5113}).

In more general case for the superpotential of such a form (where
$\alpha > 0$, $x_{0} > 0$, $n$ is natural number):
\begin{equation}
W(x) = \left\{
\begin{array}{ll}
   -\displaystyle\frac{\alpha}{|x-x_{0}|^{n}}, & \mbox{at } x<0; \\
   -\displaystyle\frac{\alpha}{|x+x_{0}|^{n}}, & \mbox{at } x>0;
\end{array} \right.
\label{eq.5.1.1.6}
\end{equation}                                          %       (5.1.1.6)
the potentials-partners $V_{1}(x)$ and $V_{2}(x)$ have the following
form:
\begin{equation}
V_{1}(x) = \left\{
\begin{array}{ll}
%   \displaystyle\frac{\alpha^{2}}{(x-x_{0})^{2n}} -
%   \displaystyle\frac{\hbar}{\sqrt{2m}}
%   \displaystyle\frac{\alpha n}{|x-x_{0}|^{n+1}} =
   \displaystyle\frac{\alpha}{(x-x_{0})^{2n}}
   \biggl(\alpha - \displaystyle\frac{\hbar n |x-x_{0}|^{n-1}}
   {\sqrt{2m}} \biggl), & \mbox{at } x<0; \\
%
%   \displaystyle\frac{\alpha^{2}}{(x+x_{0})^{2n}} -
%   \displaystyle\frac{\hbar}{\sqrt{2m}}
%   \displaystyle\frac{\alpha n}{|x+x_{0}|^{n+1}} =
   \displaystyle\frac{\alpha}{(x+x_{0})^{2n}}
   \biggl(\alpha - \displaystyle\frac{\hbar n |x+x_{0}|^{n-1}}
   {\sqrt{2m}} \biggl), & \mbox{at } x>0; \\
\end{array} \right.
\label{eq.5.1.1.7}
\end{equation}                                          %       (5.1.1.7)
\begin{equation}
V_{2}(x) = \left\{
\begin{array}{ll}
%   \displaystyle\frac{\alpha^{2}}{(x-x_{0})^{2n}} +
%   \displaystyle\frac{\hbar}{\sqrt{2m}}
%   \displaystyle\frac{\alpha n}{|x-x_{0}|^{n+1}} =
   \displaystyle\frac{\alpha}{(x-x_{0})^{2n}}
   \biggl(\alpha + \displaystyle\frac{\hbar n}{\sqrt{2m}}
   |x-x_{0}|^{n-1}\biggl), & \mbox{at } x<0; \\
%
%   \displaystyle\frac{\alpha^{2}}{(x+x_{0})^{2n}} +
%   \displaystyle\frac{\hbar}{\sqrt{2m}}
%   \displaystyle\frac{\alpha n}{|x+x_{0}|^{n+1}} =
   \displaystyle\frac{\alpha}{(x+x_{0})^{2n}}
   \biggl(\alpha + \displaystyle\frac{\hbar n}{\sqrt{2m}}
   |x+x_{0}|^{n-1}\biggl), & \mbox{at } x>0.
\end{array} \right.
\label{eq.5.1.1.8}
\end{equation}                                          %       (5.1.1.8)

The potential $V_{1}(x)$ is constant only at fulfillment of one
condition: $n=0$ or $n=1$. The condition $n=0$ gives the trivial
solution. At $n=1$ the potential $V_{2}(x)$ becomes reflectionless,
if the condition (\ref{eq.5.1.1.4}) is fulfilled. If $n \ne 1$,
then it is impossible to achieve a constancy of the potentials
$V_{1}(x)$ or $V_{2}(x)$ by variation of the parameters $\alpha$
and $m$. At replacement of the sign for $W(x)$ the sign before
the potentials $V_{1}(x)$ and $V_{2}(x)$ does not changed.
% ***************************************************************************

% ***************************************************************************
\subsubsection{Construction of the hierarchy of the reflectionless
inverse power potentials
\label{sec.5.1.2}}

One can construct hierarchy of the reflectionless potentials (where
there is tunneling), which have the inverse power dependence on
spatial coordinate. Let's consider a potential of such form:
\begin{equation}
  V_{n} (x) = \left\{
  \begin{array}{ll}
    \gamma \displaystyle\frac{\alpha^{2}}{(x-x_{0})^{2}}, &
      \mbox{at } x<0; \\
    \gamma \displaystyle\frac{\alpha^{2}}{(x+x_{0})^{2}}, &
      \mbox{at } x>0.
  \end{array} \right.
\label{eq.5.1.2.1}
\end{equation}                                          %       (5.1.2.1)
We shall find for it a potential-partner $V_{n+1}(x)$ with a
number ``up'', which has the inverse power dependence also. For
simplicity, we shall consider the potentials in the region $x > 0$
only (solutions for the potentials in the region $x < 0 $ can be
obtained by using of the replacement $x_{0} \to -x_{0}$). Write:
\begin{equation}
\begin{array}{l}
  V_{n}(x) =
        W_{n}^{2}(x) - \alpha \displaystyle\frac{d W_{n}(x)}{dx} =
        \displaystyle\frac{\gamma\alpha^{2}}{(x+x_{0})^{2}}; \\
  V_{n+1}(x) =
        W_{n}^{2}(x) + \alpha \displaystyle\frac{d W_{n}(x)}{dx}.
\end{array}
\label{eq.5.1.2.2}
\end{equation}                                          %       (5.1.2.2)
From here we obtain the equation for calculation the $W_{n}(x)$:
\begin{equation}
  \alpha \displaystyle\frac{d W_{n}(x)}{dx} =
  W_{n}^{2}(x) - 
  \displaystyle\frac{\gamma\alpha^{2}}{(x+x_{0})^{2}}.
\label{eq.5.1.2.3}
\end{equation}                                          %       (5.1.2.3)
If to find the solution in the form:
\begin{equation}
  W_{n}(x) = -\displaystyle\frac{\beta}{x+x_{0}},
\label{eq.5.1.2.4}
\end{equation}                                          %       (5.1.2.4)
%
% \begin{equation}
% \begin{array}{ll}
%   \displaystyle\frac{d W_{n}(x)}{dx} =
%     \displaystyle\frac{\beta}{(x+x_{0})^{2}}; &
%   W_{n}^{2}(x) =
%    \displaystyle\frac{\beta^{2}}{(x+x_{0})^{2}}.
% \end{array}
% \label{eq.5.1.2.5}
% \end{equation}                                          %       (5.1.2.5)
%
then we obtain the following result:
\begin{equation}
  \beta =
    \displaystyle\frac{\alpha}{2}
    \biggl(1 \pm \sqrt{4\gamma+1} \biggr).
\label{eq.5.1.2.6}
\end{equation}                                          %       (5.1.2.6)
%
% Таким образом, мы получили следующее выражение для суперпотенциала
% $W_{n}(x)$:
%
% \begin{equation}
%   W_{n}(x) =
%   \displaystyle\frac{ -\alpha (1 \pm \sqrt{4\gamma+1})}
%   {2(x+x_{0})}.
% \label{eq.5.1.2.7}
% \end{equation}                                          %       (5.1.2.7)

Now we shall find the potential-partner $V_{n+1}(x)$. According to
(\ref {eq.5.1.2.2}) and (\ref {eq.5.1.2.4}), we obtain:
\begin{equation}
  V_{n+1}(x) = 
  \biggl(1 + \gamma \pm \sqrt{4\gamma+1} \biggr)
  \displaystyle\frac{\alpha^{2}}{(x+x_{0})^{2}}.
\label{eq.5.1.2.8}
\end{equation}                                          %       (5.1.2.8)
On the basis of this expression and (\ref {eq.5.1.2.1}) one can
construct the recurrent formula for calculation of new inverse power
potential with the numbers ``up'' and ``down'' in the SUSY-hierarchy
on the basis of the known old potential (we write the potential on the
whole axis $x$):
\begin{equation}
\begin{array}{ll}
  \gamma_{n \pm 1} = 1 + \gamma_{n} \pm \sqrt{4\gamma_{n}+1}; &
  V_{n}(x) = \left\{
  \begin{array}{ll}
    \gamma_{n} \displaystyle\frac{\alpha^{2}}{(x-x_{0})^{2}}, &
      \mbox{for } x<0; \\
    \gamma_{n} \displaystyle\frac{\alpha^{2}}{(x+x_{0})^{2}}, &
      \mbox{for } x>0.
  \end{array} \right.
\end{array}
\label{eq.5.1.2.10}
\end{equation}                                          %       (5.1.2.10)

If to choose
\begin{equation}
  \gamma_{n} = 2,
\label{eq.5.1.2.11}
\end{equation}                                          %       (5.1.2.11)
than the potential (\ref{eq.5.1.2.1}) becomes reflectionless of
the form (\ref{eq.5.1.1.2}) and (\ref{eq.5.1.1.3}). In such a case
the expression (\ref{eq.5.1.2.10}) describes by the recurrent way
the sequence of the reflectionless inverse power potentials with
the following coefficients $\gamma$ (we write only first values):
\begin{equation}
  \gamma = 0, 2, 6, 12, 20, 30, 42...
\label{eq.5.1.2.12}
\end{equation}                                          %       (5.1.2.12)
and makes the SUSY-hierarchy (it is obtained for the first time).

\emph{Note, that this sequence contains natural numbers only!}
% ***************************************************************************

% ***************************************************************************
\subsubsection{A generalization on a spherically-symmetric case
\label{sec.5.1.3}}

The analysis, fulfilled above, of existence of the one-dimensional
reflectionless potentials can be generalized on the spherically
symmetric case (at $l=0$). Here, we consider the functions $W(r)$
and $V_{1,2}(r)$ for positive $r > 0$ only. For $n=1$ we obtain:
\begin{equation}
   \biggl(W(r) = \displaystyle\frac{-\alpha}{r+r_{0}}\biggr)
   \Longrightarrow
\left\{
\begin{array}{l}
   V_{1}(r) = \displaystyle\frac{\alpha}{(r+r_{0})^{2}}
        \biggl(\alpha - \displaystyle\frac{\hbar}{\sqrt{2m}}\biggl), \\
   V_{2}(r) = \displaystyle\frac{\alpha}{(r+r_{0})^{2}}
        \biggl(\alpha + \displaystyle\frac{\hbar}{\sqrt{2m}}\biggl).
\end{array} \right.
\label{eq.5.1.3.1}
\end{equation}                                          %       (5.1.3.1)
At fulfillment (\ref{eq.5.1.1.4}) the potential $V_{1}(r)$ becomes
zero, the potentials $V_{1}(r)$ and $V_{2}(r)$ --- reflectionless,
and a scattering of a particle on them --- \emph{resonant}.
% ***************************************************************************

% ***************************************************************************
\subsection{A variety of SUSY-hierarchies
\label{sec.5.2}}
% ***************************************************************************

% ***************************************************************************
\subsubsection{
A hierarchy generated by the hyperbolic superpotential
$W(x) \sim \coth{(x)}$
\label{sec.5.2.1}}

Let's consider a superpotential of a form:
\begin{equation}
  W_{n}(x) =
  |A|B \coth
  {\biggl( C \displaystyle\frac{|A|}{\alpha} (x+x_{0}) \biggr)}.
\label{eq.5.2.1.1}
\end{equation}                                          %       (5.2.1.1)
We find potentials-partners connected by this superpotential:
\begin{equation}
\begin{array}{lcl}
  V_{n}(x) & = &
    |A|^{2} \left( B^{2} + \displaystyle\frac{B^{2} + BC}
    {\sinh^{2}\biggl( C \displaystyle\frac{|A|}{\alpha}
               (x+x_{0}) \biggr)} \right), \\
  V_{n+1}(x) & = &
    |A|^{2} \left( B^{2} + \displaystyle\frac{B^{2} - BC}
    {\sinh^{2}\biggl( C \displaystyle\frac{|A|}{\alpha}
               (x+x_{0}) \biggr)} \right).
\end{array}
\label{eq.5.2.1.2}
\end{equation}                                          %       (5.2.1.2)
We see, that the first item in the potentials-partners $V_{n}(x)$
and $V_{n+1}(x)$, connected through the superpotential of the form
(\ref{eq.5.2.1.1}), does not changed. Then the following potentials
in this hierarchy have the same first item. The constant potential
can belong to such hierarchy only, when:
\begin{equation}
  B^{2} = 1.
\label{eq.5.2.1.3}
\end{equation}                                          %       (5.2.1.3)
One can use:
\begin{equation}
\begin{array}{ll}
  B = -1; & C > 0.
\end{array}
\label{eq.5.2.1.4}
\end{equation}                                          %       (5.2.1.4)

If to enter designations:
\begin{equation}
\begin{array}{ll}
  \beta_{n} = 1-C, & \beta_{n+1} = 1+C,
\end{array}
\label{eq.5.2.1.6}
\end{equation}                                          %       (5.2.1.6)
then we obtain a general form of the potential belonging to the
SUSY-hierarchy of potentials, connected through the superpotential
(\ref{eq.5.2.1.1}):
\begin{equation}
\begin{array}{ll}
  V_{n}(x) =
    |A|^{2} \left( 1 + \displaystyle\frac{\beta_{n}}
    {\sinh^{2}\biggl(C \displaystyle\frac{|A|}{\alpha}
               (x+x_{0}) \biggr)} \right); &
  \beta_{n+1} = 2-\beta_{n}.
\end{array}
\label{eq.5.2.1.8}
\end{equation}                                          %       (5.2.1.8)

One can find an useful interdependence between the
potentials-partners:
\begin{equation}
  V_{n+1}(x) =
    -V_{n}(x) + |A|^{2} \left( 2 +
      \displaystyle\frac{2}
      {\sinh^{2}\biggl( C \displaystyle\frac{|A|}{\alpha}
               (x+x_{0}) \biggr)} \right).
\label{eq.5.2.1.10}
\end{equation}                                          %       (5.2.1.10)

If to take into account, that the constant potential (\ref{eq.3.1})
belongs to such hierarchy, than from (\ref{eq.5.2.1.10}) one can
obtain its SUSY-partner coincided with the third solution of the
system (\ref{eq.4.1.22}) at $C=1$. Thus, the SUSY-hierarchy,
obtained by such a way, contains only two potentials, which are
reflectionless:
\begin{equation}
\begin{array}{lcl}
  V_{n}(x) & = & |A|^{2}; \\
  V_{n\pm 1}(x) & = &
    |A|^{2} \left( 1 + \displaystyle\frac{2}
      {\sinh^{2}\biggl( C \displaystyle\frac{|A|}{\alpha}
               (x+x_{0}) \biggr)} \right).
\end{array}
\label{eq.5.2.1.11}
\end{equation}                                          %       (5.2.1.11)
One can see that there are no other potentials in this
SUSY-hierarchy, with the exception of these two. The expression
(\ref{eq.5.2.1.10}) corresponds to the definition of the shape
invariant potentials (see~\cite{Cooper.1995.PRPLC}, p.~290, (96)).

On the other side, from (\ref{eq.5.2.1.8}), (\ref{eq.5.2.1.4}) and
(\ref{eq.5.2.1.6})) one can find a sequence of values of $\beta_{n}$
and $C$:
\begin{equation}
\begin{array}{lcl}
  \biggl(
  (\beta_{n} = 0) \to
  (\beta_{n+1} = 2) \to
  (\beta_{n+2} = 0) \biggr)
  & \Longrightarrow &
  (\beta_{i} = 0, 2; \, C=1).
\end{array}
\label{eq.5.2.1.12}
\end{equation}                                          %       (5.2.1.12)
We again come to the conclusion about the existence of only two
reflectionless potentials in the considered hierarchy.

{\bf Property:} if we add the following potential
\[
% \begin{equation}
  V(x) =
    \displaystyle\frac{2 |A|^{2}}
    {\sinh^{2}\biggl( \displaystyle\frac{|A|}{\alpha}
    (x+x_{0}) \biggr)},
% \label{eq.5.2.1.13}
% \end{equation}                                          %       (5.2.1.13)
\]
to the constant potential $V(x) = |A|^{2}$, then new potential
remains reflectionless.
% ***************************************************************************

% ***************************************************************************
\subsubsection{
A hierarchy generated by the hyperbolic superpotential
$W(x) \sim \tanh{(x)}$
\label{sec.5.2.2}}

Let's find the potentials-partners connected by a superpotential of
such a form:
\begin{equation}
  W_{n}(x) =
  |A|B \tanh
  \biggl( C \displaystyle\frac{|A|}{\alpha} (x+x_{0}) \biggr).
\label{eq.5.2.2.1}
\end{equation}                                          %       (5.2.2.1)
We obtain:
\begin{equation}
\begin{array}{lcl}
  V_{n}(x) & = &
    |A|^{2} \left( B^{2} - \displaystyle\frac{B^{2} + BC}
    {\cosh^{2}\biggl( C \displaystyle\frac{|A|}{\alpha}
               (x+x_{0}) \biggr)} \right), \\
  V_{n+1}(x) & = &
    |A|^{2} \left( B^{2} - \displaystyle\frac{B^{2} - BC}
    {\cosh^{2}\biggl( C \displaystyle\frac{|A|}{\alpha}
               (x+x_{0}) \biggr)} \right).
\end{array}
\label{eq.5.2.2.2}
\end{equation}                                          %       (5.2.2.2)

Similarly the reasoning of the previous paragraph, we obtain:
\begin{equation}
\begin{array}{ll}
  B = -1; & C > 0.
\end{array}
\label{eq.5.2.2.3}
\end{equation}                                          %       (5.2.2.3)

If to enter the following designations:
\begin{equation}
\begin{array}{ll}
  \beta_{n} = 1-C, & \beta_{n+1} = 1+C,
\end{array}
\label{eq.5.2.2.5}
\end{equation}                                          %       (5.2.2.5)
then we obtain recurrent dependences for description of a general
form of the potential belonging to the SUSY-hierarchy with the
superpotential (\ref{eq.5.2.2.1}):
\begin{equation}
\begin{array}{lcl}
  V_{n}(x) & = &
    |A|^{2} \left( 1 - \displaystyle\frac{\beta_{n}}
    {\cosh^{2}\biggl( C \displaystyle\frac{|A|}{\alpha}
               (x+x_{0}) \biggr)} \right), \\
  \beta_{n+1} & = & 2-\beta_{n},
\end{array}
\label{eq.5.2.2.7}
\end{equation}                                          %       (5.2.2.7)
and interdependence between the potentials-partners in this
hierarchy:
\begin{equation}
  V_{n+1}(x) =
    -V_{n}(x) + |A|^{2} \left( 2 -
      \displaystyle\frac{2}
      {\cosh^{2}\biggl( C \displaystyle\frac{|A|}{\alpha}
               (x+x_{0}) \biggr)} \right).
\label{eq.5.2.2.9}
\end{equation}                                          %       (5.2.2.9)

If as one of such potentials to use the constant potential
(\ref{eq.3.1}), then one can conclude that the SUSY-hierarchy,
defined by such a way, contains only two reflectionless potentials
of the form:
\begin{equation}
\begin{array}{lcl}
  V_{n}(x) & = & |A|^{2}; \\
  V_{n\pm 1}(x) & = &
    |A|^{2} \left( 1 - \displaystyle\frac{2}
      {\cosh^{2}\biggl(\displaystyle\frac{|A|}{\alpha}
               (x+x_{0}) \biggr)} \right).
\end{array}
\label{eq.5.2.2.10}
\end{equation}                                          %       (5.2.2.10)

One can find all possible values of the coefficient $\beta_{n}$:
\begin{equation}
\begin{array}{ll}
  \beta_{i} = 0, 2; &
  C = 1.
\end{array}
\label{eq.5.2.2.11}
\end{equation}                                          %       (5.2.2.11)

{\bf Property:} if to subtract the potential of the form
\[
% \begin{equation}
  V(x) =
    \displaystyle\frac{2 |A|^{2}}
    {\cosh^{2}\biggl( \displaystyle\frac{|A|}{\alpha}
    (x+x_{0}) \biggr)},
\label{eq.5.2.2.12}
% \end{equation}                                          %       (5.2.2.12)
\]
from the constant potential $V(x) = |A|^{2}$, then new potential
remains reflectionless.
% ***************************************************************************

% ***************************************************************************
\subsection{Methods
\label{sec.5.3}}

Now we shall be looking for an approach for calculation of all
possible forms of potentials belonging to one SUSY-hierarchy, which
contains a reflectionless potential $V_{n}(x)$. Let's consider a
sequence of the potentials of such hierarchy in a direction of
numbers ``up''. According to (\ref{eq.2.3}), we write:
\begin{equation}
  V_{n+1} (x) = 
  W_{n}^{2}(x) + \alpha \displaystyle\frac{d W_{n}(x)}{dx} =
  V_{n}(x) + 2 \alpha \displaystyle\frac{d W_{n}(x)}{dx}.
\label{eq.5.3.1}
\end{equation}                                          %       (5.3.1)
For the following numbers $n+1$... $n+m$ we obtain:
\begin{equation}
  V_{n+m} (x) = 
  V_{n}(x) + 2 \alpha \biggl( \displaystyle\frac{d W_{n}(x)}{dx} +
    \displaystyle\frac{d W_{n+1}(x)}{dx} + ... +
    \displaystyle\frac{d W_{n+m-1}(x)}{dx} \biggr).
\label{eq.5.3.3}
\end{equation}                                          %       (5.3.3)
% ***************************************************************************

% ***************************************************************************
\subsubsection{A recurrent method of construction of the
SUSY-hierarchy of the reflectionless potentials
\label{sec.5.3.1}}

The superpotential $W_{n+1}(x)$ can be expressed through $W_{n}(x)$:
\begin{equation}
  W_{n+1}^{2}(x) - \alpha \displaystyle\frac{d W_{n+1}(x)}{dx} =
  W_{n}^{2}(x) + \alpha \displaystyle\frac{d W_{n}(x)}{dx}.
% = V_{n+1}(x).
\label{eq.5.3.1.1}
\end{equation}                                          %       (5.3.1.1)
Using the function $W_{n}(x)$ with the number $n$ and solving the
differential equation (\ref{eq.5.3.1.1}), one can find the function
$W_{n+1}(x)$ with the number $n+1$. From the equation
(\ref{eq.5.3.1.1}) one can see, that the function $W_{n+1}(x)$ of the
form
\begin{equation}
  W_{n+1}(x) =  -W_{n}(x)
\label{eq.5.3.1.2}
\end{equation}                                          %       (5.3.1.2)
is a partial solution of this equation (in particular, this
expression is carried out for the superpotentials (\ref{eq.4.1.17})
and (\ref{eq.4.2.4}) with numbers ``up'' and ``down'' concerning
the constant potential).
Knowing the partial solution decision for the function $W_{n+1}(x)$,
one can find its general solution by a method described in the
following subsection. Further, knowing the general form of the
function $W_{n+1}(x)$, one can construct a new equation for
determination of an unknown function $W_{n+2}(x)$. So, we obtain
the recurrent approach for calculation of the general form of the
function $W_{i}(x)$ with numbers from $n$ up to $n+m$ (i.~e. we
take into account all possible solutions). The determined
superpotentials $W_{i}(x)$ allow to find potentials $V_{i}(x)$ with
numbers from $n$ up to $n+m$:
\begin{equation}
  V_{i}(x) =
  W_{i}^{2}(x) - \alpha \displaystyle\frac{d W_{i}(x)}{dx}
\label{eq.5.3.1.3}
\end{equation}                                          %       (5.3.1.3)
By such a way, one can construct a general form of the SUSY-hierarchy
(it is necessary to take into account potentials with numbers
``down''), if we know its one potential $V_{n}(x)$.

In order to construct the SUSY-hierarchy of the reflectionless
potentials, it is need to use the constant potential (\ref{eq.3.1})
as the potential $V_{n}(x)$ and on the basis of the described above
approach consistently to determine possible types of the potentials.
Here, knowledge of wave functions of these potentials is not
required. Further, from this set of the potentials it is need to
select the reflectionless potentials, which are finite in whole
axis $x$ and converge to uniquely finite limits in the asymptotic
areas.
% ***************************************************************************

% ***************************************************************************
\subsubsection{Search of a general form of the superpotential
\label{sec.5.3.2}}

We shall find a general solution of the equation (\ref{eq.5.3.1.1}).
Let's rewrite this equation by such a way:
\begin{equation}
  \displaystyle\frac{d W(x)}{dx} -
  \displaystyle\frac{1}{\alpha} W^{2}(x) =
  - \displaystyle\frac{F(x)}{\alpha}.
\label{eq.5.3.2.1}
\end{equation}                                          %       (5.3.2.1)
This is \emph{the Ricatti equation}. According to
\cite{Tihonov.1998} (see~p.~29), it is not integrated exactly.
However, it has one property: \emph{if we know its one partial
solution, then this equation can be reduced to the Bernoulli
equation and one can find its general solution}.
Let we know its partial solution $\tilde{W}(x)$ (for example,
see~(\ref{eq.5.3.1.2})). We change variable:
\begin{equation}
  z(x) = W(x) - \tilde{W}(x).
\label{eq.5.3.2.2}
\end{equation}                                          %       (5.3.2.2)
Then one can transform the equation (\ref{eq.5.3.2.1}) into the
following equation with new variables:
\begin{equation}
  \displaystyle\frac{dz(x)}{dx} -
  \displaystyle\frac{2 \tilde{W}(x)}{\alpha} z(x) -
  \displaystyle\frac{1}{\alpha} z^{2}(x) = 0.
\label{eq.5.3.2.3}
\end{equation}                                          %       (5.3.2.3)
This is the Bernoulli equation. We change variable:
\begin{equation}
  y(x) = z^{-1} = \displaystyle\frac{1}{z(x)}
\label{eq.5.3.2.4}
\end{equation}                                          %       (5.3.2.4)
and reduce this equation to such a form:
\begin{equation}
  \displaystyle\frac{dy(x)}{dx} +
  \displaystyle\frac{2\tilde{W}(x)}{\alpha} y(x) =
  - \displaystyle\frac{1}{\alpha}.
\label{eq.5.3.2.5}
\end{equation}                                          %       (5.3.2.5)

In result, we obtain the general solution:
\begin{equation}
  y(x) =
  e^{-\displaystyle\int \displaystyle\frac{2\tilde{W}(x)}{\alpha} dx}
  \biggl( C_{1} -
  \displaystyle\frac{1}{\alpha}
  \displaystyle\int
  e^{\displaystyle\int
     \displaystyle\frac{2\tilde{W}(x)}{\alpha} dx} dx \biggl),
\label{eq.5.3.2.6}
\end{equation}                                          %       (5.3.2.6)
where $C_{1}$ is a constant of integration (the indefinite integral
is used). In the old variables the solution has the following form:
\begin{equation}
\begin{array}{lcl}
  W(x) & = &
  z(x) + \tilde{W}(x) =
  \displaystyle\frac{1}{y(x)} + \tilde{W}(x) =
  \displaystyle\frac{
  e^{\displaystyle\int
    \displaystyle\frac{2\tilde{W}(x)}{\alpha} dx}}
  {C_{1} -
    \displaystyle\frac{1}{\alpha}
     \displaystyle\int
     e^{\displaystyle\int
     \displaystyle\frac{2\tilde{W}(x)}{\alpha} dx} dx} + \\
  & + & \tilde{W}(x).
\end{array}
\label{eq.5.3.2.7}
\end{equation}                                          %       (5.3.2.7)

Using the different partial solutions $\tilde{W}(x)$, one can find
new types of the reflectionless potentials, which have a simple
analytical form. For the selected function $\tilde{W}(x)$ one can
obtain the general form of the superpotential $W(x)$ and the
potential-partner, which depend on the constant $C_{1}$. The
constant $C _ {1} $ plays a role of a free parameter, changing
which, one can change the form of the superpotential and the
potential-partner (such potentials with different forms will make
the one-parameter family of isospectral potentials with the
parameter $C_{1}$). In a general case, the change of the potential
form is not reduced to its displacement along the axis $x$ as in the
case of the displacement of the potentials (\ref{eq.4.1.22}) and
(\ref{eq.4.2.5}) at the change of the parameter $x_{0}$. If to
choose any form of the functions (\ref{eq.4.1.17}) as the partial
solution $\tilde{W}(x)$ and to put $C_{1}=0$, then on a basis
(\ref{eq.5.3.2.7}) one can obtain solutions (\ref{eq.4.1.22})
already known. But if to change the parameter $C_{1}$, than we
obtain new reflectionless potentials.

For example, let's consider the found earlier family of the inverse
power reflectionless potentials (\ref{eq.5.1.2.10}). We select from
it one a potential with a number $n$. We assume that we know a form
of this potential and we shall search for it all possible
potentials-partners (with the number $n+1$).
Here, the definition (\ref{eq.5.1.2.10}) makes the potential with
the number $n$ and all its SUSY potentials-partners spatially
symmetric at change $x \to -x$.
According to subsection~\ref{sec.5.1.2}, we have one known partial
solution for the superpotential $W_{n}(x)$, which connects the
potentials $V_{n}(x)$ and $V_{n+1}(x)$ among themselves, and has the
form (\ref{eq.5.1.2.4}) with taking into account (\ref{eq.5.1.2.6}),
where $\gamma_{n}$ must be chosen from the sequence
(\ref{eq.5.1.2.12}). Then on the basis of (\ref{eq.5.3.2.7}) we
find the general form of the superpotential:
\begin{equation}
  W(x) =
    \left \{
    \begin{array}{cl}
      \displaystyle\frac{2\beta - \alpha}
        {C_{1}(2\beta-\alpha)
        x^{2\beta / \alpha} + x} -
        \displaystyle\frac{\beta}{x}, &
        \mbox{при } 2\beta \ne \alpha; \\
      \displaystyle\frac{\alpha}{x}
        \biggl(
        \displaystyle\frac{1}{\alpha C_{1} - \log{x}} -
        \displaystyle\frac{1}{2} \biggr), &
        \mbox{при } 2\beta = \alpha.
    \end{array} \right.
\label{eq.5.3.2.8}
\end{equation}                                          %       (5.3.2.8)
On the basis of the second equation of the system (\ref{eq.5.1.2.2})
one can calculate a general form of the potential-partner
$V_{n+1}(x)$. Further, on the basis of the first equation of the
system (\ref{eq.5.1.2.2}) and already chosen form of the potential
$V_{n+1}(x)$ one can find possible variations of the potential
$V_{n}(x)$ (which remains reflectionless).

\begin{figure}[t]
\includegraphics[width=8cm]{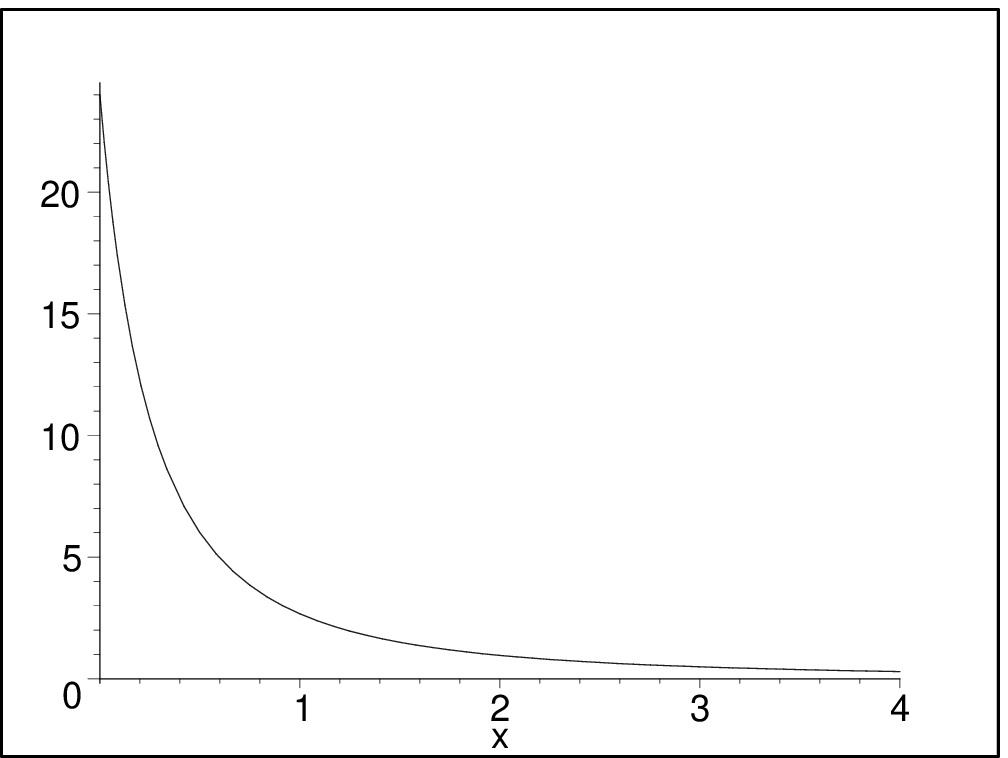}
\includegraphics[width=8cm]{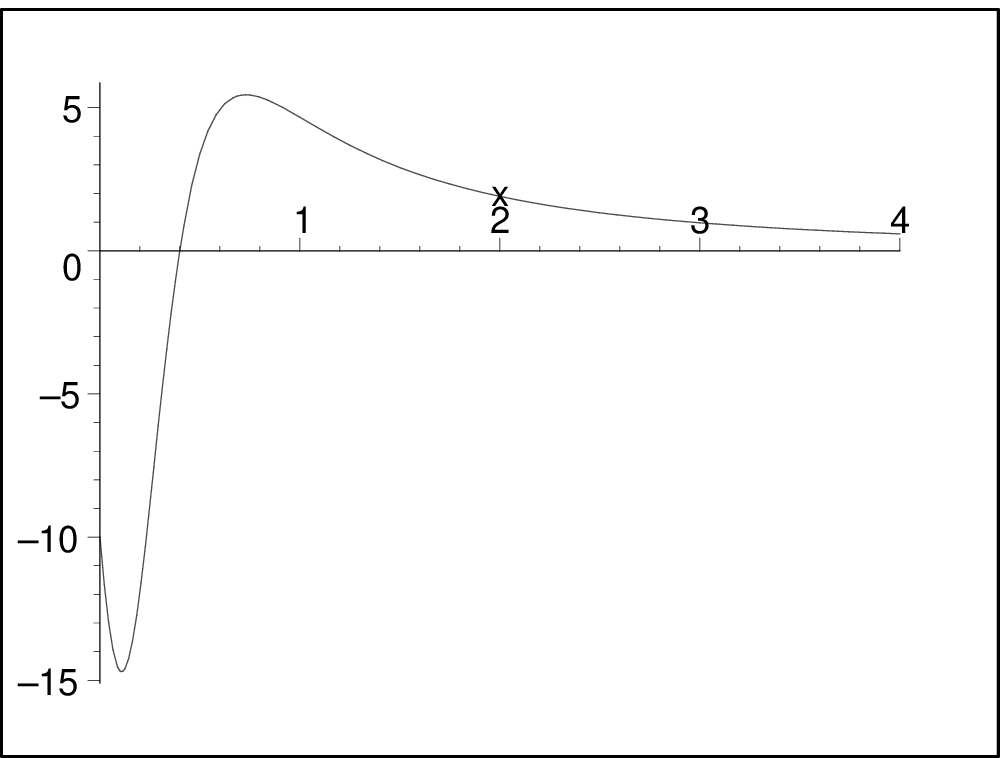}
\caption{Reflectionless potentials $V_{n}(x)$ and $V_{n+1}(x)$ for
values $C_{1}=1$, $\gamma=6$ and $x_{0}=0.5$
\label{fig.5321}}
\end{figure}

\begin{figure}[t]
\includegraphics[width=8cm]{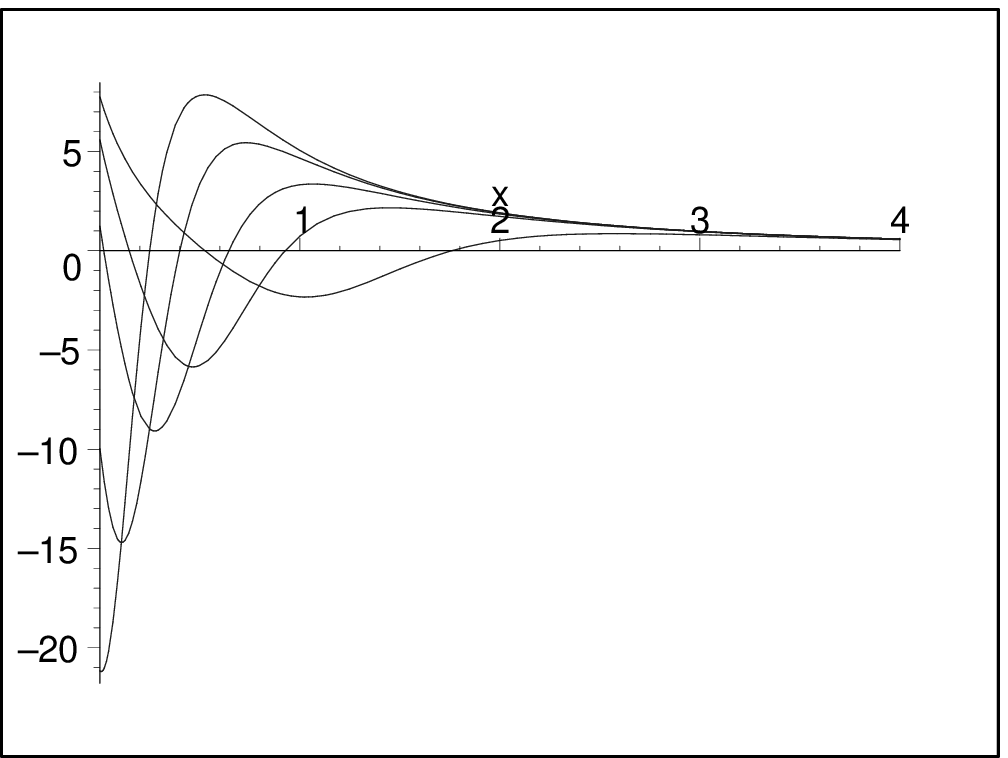}
\includegraphics[width=8cm]{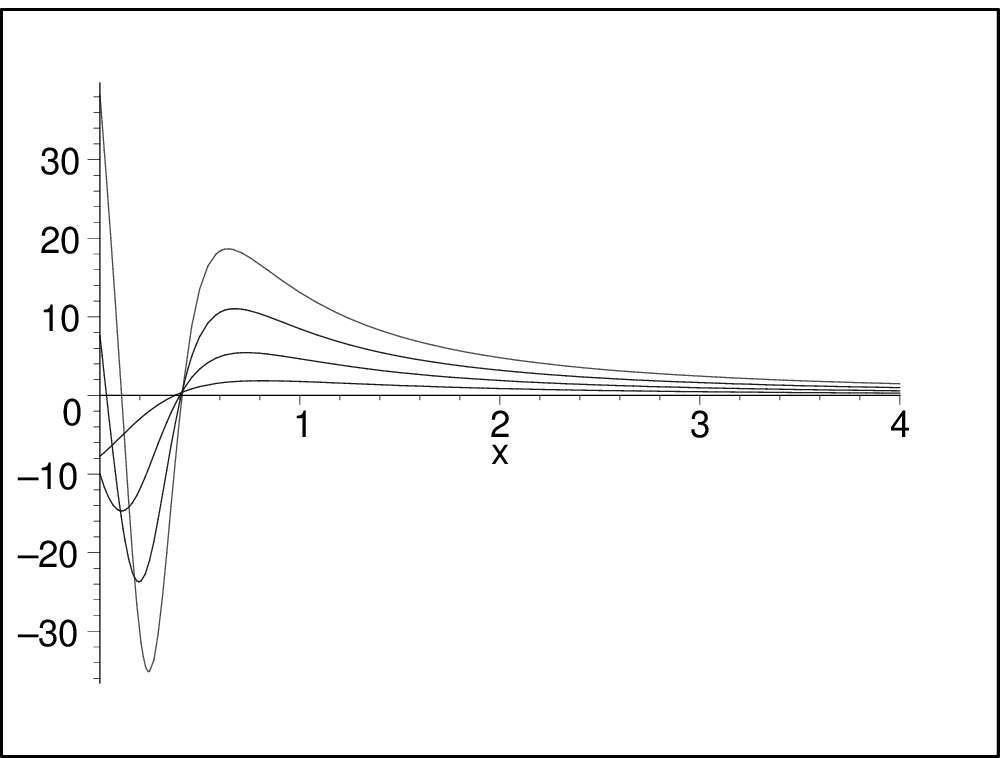}
\caption{
A dependence of the reflectionless potential $V_{n+1}(x)$ on the
parameters $C_{1}$ and $\gamma$:
at the change of the parameter $C_{1}$ the barrier maximum and
the hole minimum of the potential is displaced along the axis $x$
(the first figure, the values $C_{1} = 0.01, 0.1, 0.3, 1.0, 2.5$,
$\gamma=6$, $x_{0}=0.5$),
at the change of the parameter $\gamma$ the barrier maximum of the
potential practically does not displaced along the axis $x$
(the second figure, values $C_{1} = 1$, $\gamma=2, 6, 12, 20 $,
$x_{0}=0.5$)
\label{fig.5322}}
\end{figure}

\begin{figure}[t]
\includegraphics[width=8cm]{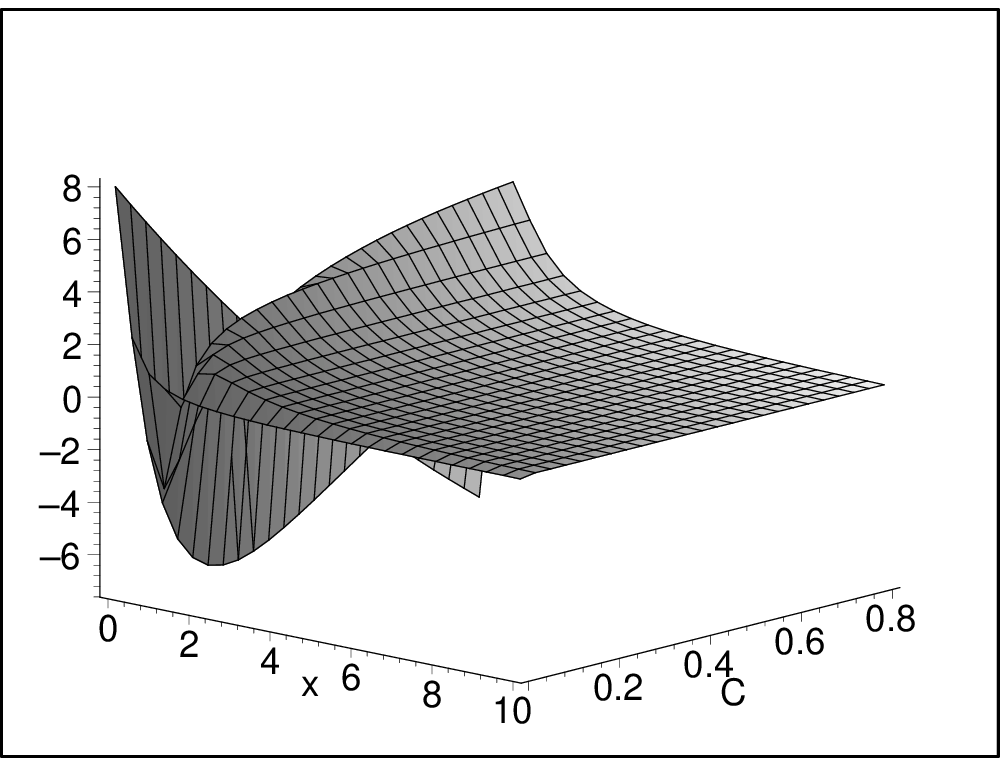}
\includegraphics[width=8cm]{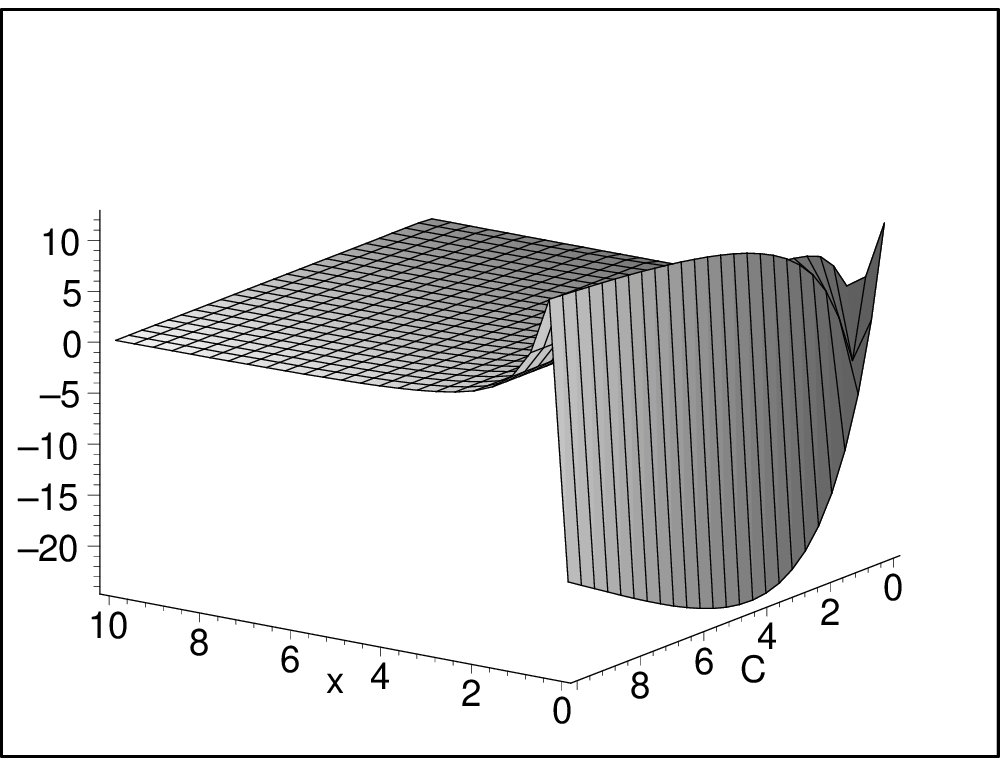}
\caption{Continuous change of the form of the reflectionless
potential $V_{n+1}(x)$ (with the barrier and the hole) at
variation of the parameter $C_{1}$ ($\gamma=6$, $x_{0}=0.5$)
\label{fig.5323}}
\end{figure}

\begin{figure}[t]
\includegraphics[width=8cm]{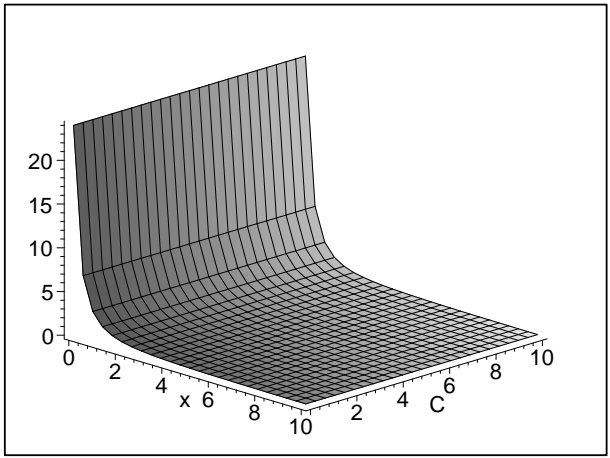}
\caption{The reflectionless potential $V_{n+1}(x)$ does not
changed at the change of the parameters $C_{1}$ and $\gamma$
(the diagrams coincide for different values $\gamma=2, 6, 12, 20$,
$x_{0}=0.5$)
\label{fig.5324}}
\end{figure}

Let's analyze the change of the form of the potentials $V_{n}(x)$
and $V_{n+1}(x)$ in variation of the parameters $C_{1}$ and $\gamma$
in the case $2\beta \ne \alpha$. Diagrams of these potentials for
values $C_{1}=1$ and $\gamma=6$ on the positive semi-axis $x \ge 0$
are shown in Fig.~\ref{fig.5321}. From here one can see, that
the potential $V_{n}(x)$ remains inverse power, but a barrier and a
hole appear in the potential $V_{n+1}(x)$. This second potential is
qualitatively similar to a radial interaction potential in the
elastic scattering of particles on nuclei and in decay of compound
nuclei in the spherically symmetric consideration. Such a form of
the reflectionless potential and its application to the scattering
theory are found for the first time.
From Fig.~\ref{fig.5322} one can see, that locations of the barrier
maximum and the hole minimum of the potential $V_{n+1}(x)$ are
shifted along the axis $x$ at the change of the parameter $C_{1}$,
but the change of the parameter $\gamma$ practically does not
influence on the location of the barrier maximum of this potential.
In Fig.~\ref{fig.5323} we show the continuous change of the form of
this potential at the change of the parameter $C_{1}$ (here one can
see, how the barrier and hole are changed).
In accordance with preliminary calculations, the change of the
parameters does not influence on the form of the potential
$V_{n}(x)$ (see~Fig.~\ref{fig.5324}).

From (\ref{eq.5.3.2.8}) one can see, that at $C_{1} \to 0$ (at
$2\beta \ne \alpha$) the superpotential tends to the earlier known
function (\ref{eq.5.1.2.4}) with taking into account of
(\ref{eq.5.1.2.6}). But at $C_{1} \ne 0$ we obtain new
reflectionless potentials (and the superpotentials, corresponding
to them).
% ***************************************************************************

% ***************************************************************************
\subsubsection{Scaling of the reflectionless potential
\label{sec.5.3.3}}

If we know the reflectionless potential $V_{n}(x)$, then the
potential $V_{n+1}(x) = C V_{n}(x)$ (where $C=const$) is also
reflectionless. Really, let $V_{n}(x)$ is the reflectionless
potential. Then we write:
\begin{equation}
\begin{array}{c}
    -\displaystyle\frac{\hbar^{2}}{2m}
     \displaystyle\frac{d^{2}}{dx^{2}}
     \varphi^{(n)}(x) +
     (V_{n}(x) - E^{(n)}) \varphi^{(n)}(x) = 0, \\
     V_{n+1}(x) = C V_{n}(x), \\
     E^{(n+1)} = C E^{(n)}.
\end{array}
\label{eq.5.3.3.1}
\end{equation}                                          %       (5.3.3.1)
From here we obtain the Schr\"{o}dinger equation for the new
potential:
\begin{equation}
\begin{array}{ll}
  -\displaystyle\frac{\hbar^{2}}{2m}
  \displaystyle\frac{d^{2}}
  {d \biggl( \displaystyle\frac{x}{\sqrt{C}} \biggr)^{2}}
  \varphi^{(n)}(x) +
  (V_{n+1}(x) - E^{(n+1)}) \varphi^{(n)}(x) = 0. &
\end{array}
\label{eq.5.3.3.2}
\end{equation}                                          %       (5.3.3.2)

{\bf Conclusion:} if $V_{n}(x)$ is a reflectionless potential,
then $V_{n+1}(x) = C V_{n}(x)$ is also the reflectionless potential
having the energy spectrum $E^{(n+1)}$, compressed relatively
$E^{(n)}$, and the wave function $\varphi^{(n+1)}(x)$, compressed
along the axis $x$ relatively the wave function $\varphi^{(n)}(x)$:
\begin{equation}
\begin{array}{ll}
  E^{(n+1)} = C E^{(n)}, &
  \varphi^{(n+1)}(x) =
  \varphi^{(n)}\biggl(\displaystyle\frac{x}{\sqrt{C}} \biggr).
\end{array}
\label{eq.5.3.3.3}
\end{equation}                                          %       (5.3.3.3)
% ***************************************************************************

% ***************************************************************************
\subsubsection{Spatial symmetry of the reflectionless potentials
\label{sec.5.3.4}}

Let's analyze, whether the reflectionless potentials have a property
of the spatial symmetry or antisymmetry at change $x \to -x$.
At first we shall consider the potentials-partners to the constant
potential (\ref{eq.3.1}) with the number ``up'' and the
superpotentials, corresponding to them, in the case $x_{0} \to 0$
(i.~e. at the exception of the parameter $x_{0}$). From the general
form of the solutions from the second up to the fifth one of the
superpotential (\ref{eq.4.1.17}) and the potential-partners
(\ref{eq.4.1.22}) one can conclude:
\begin{equation}
\begin{array}{ll}
  W_{n+1}(-x) = -W_{n+1}(x),
    & \mbox{for } |W(x)| \ne |A|, x \ne 0 \mbox{ and } x_{0}=0;\\
  V_{n+1}(-x) = V_{n+1}(x),
    & \mbox{for } |W(x)| \ne |A|, x \ne 0 \mbox{ and } x_{0}=0;.
\end{array}
\label{eq.5.3.4.1}
\end{equation}                                          %       (5.3.4.1)
This property is fulfilled for the superpotential and the
potential-partner to the constant potential with the number
``down'' (see~(\ref{eq.4.2.4}) and (\ref{eq.4.2.5})). Therefore,
any potential-partner to the constant potential is spatially
symmetric at $x_{0} \to 0$ (i.~e. it keeps a sign at the change
$x \to -x$), and the superpotential, connected to it, spatially
antisymmetric (i.~e. it changes a sign at the change $x \to -x$).

Let we know a general form of the superpotential $W_{n}(x)$ with
the number $n$, which is antisymmetric function. According to
(\ref{eq.5.3.1.2}), there is a partial solution for the
superpotential with the number $n+1$, which also is antisymmetric
function. In order to analyze, whether there a general solution
for the superpotential $W_{n}(x)$ is the antisymmetric function,
we shall consider the expression (\ref{eq.5.3.2.7}). The integral
$\int\exp{(2\tilde{W}(x) /\alpha)} \, dx$ is the indefinite
integral from the antisymmetric function and, therefore, it is
the symmetric function. Then the function $W(x)$ defined by
(\ref{eq.5.3.2.7}), is the antisymmetric function at $C_{1} = 0$
also.

Thus, we have proved the following {\bf property}: any
reflectionless potential, which belongs to SUSY-hierarchy with
the constant potential, and is defined with all parameters (the
constant of integration), which are equal to zero, is spatially
symmetric, and the superpotential, concerned with it, is
spatially antisymmetric.

\emph{If one of free parameters is distinct from zero, then the
spatial symmetry of the potential can be broken, that will result
in occurrence of new spatially asymmetric reflectionless
potentials} (it is found for the first time).
One can use this idea for construction of new reflectionless
spatially asymmetrical potentials (in specific case, which are
defined by one continuous function on the whole axis $x$).
Generally, the potential of the reflectionless SUSY-hierarchy
can not be antisymmetric function (and any of all superpotentials
connecting it with the constant potential in this SUSY-hierarchy,
can not be symmetric function, with the exception for constant).

It will be interesting to apply this analysis for the known
reflectionless soliton-like potentials, reflectionless one and
many-steps shape invariant potentials (with different types of
parameter transformations), and reflectionless potentials found
on the basis of methods of inverse problem.
% ***************************************************************************

% ***************************************************************************
\section{Conclusions
\label{sec.6}}

In this paper the class of the one-dimensional reflectionless
potentials constituted into one SUSY-hierarchy is studied. Note
the following conclusions.
\begin{itemize}
\item
The new approach for determination of the general form of the
reflectionless potential on the basis of construction of the
SUSY-hierarchy, which contains the constant potential, and
extraction from it the potentials, which are finite on the whole
axis $x$ and in the asymptotic areas tend to uniquely finite
limits, is proposed.

\item
The assumption about that a reason of the absolute transparency
of any reflectionless potential consists in its SUSY-interrelation
with the constant potential, is putted.

\item
The general integrated form of the interdependence between the
superpotentials with neighboring numbers in the SUSY-hierarchy of
the reflectionless potentials (in analytical solution of the
Ricatti equation) is found.

\item
The way of construction of the new reflectionless potentials of a
simple analytical form, which are expressed through finite number
of the elementary functions on the basis of the obtained above
interdependence between the superpotentials, is pointed out.

\item
It is shown, that the constant potential has only five different
types of its SUSY potentials-partners expressed through the
elementary functions, from which the reflectionless potentials can
be selected.

\item
The analysis of existence of an absolute transparency for the
potential having the inverse power dependence on spatial
coordinate (and where tunneling is possible), i.~e. of the form
$V(x) = \pm\alpha / |x-x_{0}|^{n}$ (where $\alpha$ and $x_{0}$ are
constant, $n$ is natural number), is fulfilled. It is shown, that
such potential can be reflectionless at $n = 2$ only. The
SUSY-hierarchy of the inverse power reflectionless potentials is
constructed.

\item
The assumption about that any reflectionless potential can be
presented in an integrated form (or in a simple analytical form)
with use of the finite number of the elementary functions
(in particular, such a representation can exist for the known
reflectionless soliton-like potentials or the known shape invariant
potentials with one or many steps scaling of parameter, which are
expressed through series), is formulated.

\item
The new types of the reflectionless potentials of a simple
analytical form, which look (on the semi-axis $x$) like the
radial interaction potential with the barrier and the hole
between particles with nuclei at their elastic scattering (or for
decay of the compound spherical systems), are opened.
Such potentials can present new exactly solvable models.

\item
The new way of construction of the spatially asymmetrical
reflectionless potentials (relatively a point $x=0$) is pointed out.
\end{itemize}

Perspectives of such representations of the reflectionless
potentials consist in their simple analytical form, that is
convenient at a qualitative analysis of properties of tunneling
processes.
% ***************************************************************************

% ***************************************************************************
% \bibliographystyle{h-physrev4}
\bibliographystyle{h-elsevier3}
\bibliography{Ref}
% ***************************************************************************

% ***************************************************************************
% \newpage
% \listoffigures

\newpage

%---------------------------------------------------------------------------

%---------------------------------------------------------------------------

% ***************************************************************************

% ***************************************************************************
% \newpage
% \begin{center}
% {\large \bf Адреса:}
% \end{center}
% 
% {\bf С.~П.~Майданюк} \\
% 
% Iнститут ядерних дослiджень НАН Укра∙ни, \\
% просп. Науки, 47, Ки∙в-28, 03680, Укра∙на.                     \\
% E-mail: {\em maidan@kinr.kiev.ua, Sergei.Maydanyuk@fuw.edu.pl} \\
% тел.:   {\em (8-044-) 265-46-92.}
\end{document}